\documentclass[conference]{IEEEtran} 

\usepackage{_cfg-paper}

\newboolean{makeXTRA}
\setboolean{makeXTRA}{false} 

\usepackage{_cfg-me}

\ifthenelse{\boolean{makeXTRA}}
  {\newcommand{\XTRA}[1]{\phantom{}\begingroup\slshape\color{RoyalBlue}\ignorespaces#1\ignorespaces\endgroup}}
  {\newcommand{\XTRA}[1]{}}

\newcommand*{\circleb}[1]{\tikz[baseline=(char.base)]{
            \node[shape=circle,fill,inner sep=0.5pt] (char) {\textcolor{white}{#1}};}}

\newcounter{obsc}
\newcommand*{\obs}[1]{%
    \stepcounter{obsc}%
    {\textbf{Observation ~\arabic{obsc}}}%
    }
\definecolor{bg}{rgb}{0.97,0.97,0.97}
\usepackage{pifont}   
\usepackage{float}    
\usepackage{xcolor}   

\newcommand{\cmark}{\textcolor{blue}{\ding{51}}} 
\newcommand{\xmark}{\textcolor{red}{\ding{55}}}   
  
\newlist{myChallenge}{enumerate}{1}
\setlist[myChallenge]{label=C$_{\arabic*}$}

\usepackage[capitalize,noabbrev]{cleveref}
\Crefname{section}{Section}{Sections}
\Crefformat{section}{#2#1#3}
\crefname{section}{\S}{\S}
\crefformat{section}{#2\S#1#3}

\Crefname{figure}{Figure}{Figures}
\crefname{figure}{Fig.}{Figs.}

\Crefname{equation}{Equation}{Equations}
\crefname{equation}{Eq.}{Eqs.}

\crefname{algocf}{Alg.}{Algs.}    
\Crefname{algocf}{Algorithm}{Algorithms}

\definecolor{shadecolor}{gray}{0.9} 

\usepackage{bbding}
\newcommand{\COMPRESS}[1]{}

\providecommand{\linesref}[2]{\hyperref[#1]{Lines~\ref*{#1}--\ref*{#2}}}
\providecommand{\lineref}[1]{\hyperref[#1]{Line~\ref*{#1}}}

\usepackage[ruled,vlined,linesnumbered]{algorithm2e}

\usepackage{float}
\floatstyle{ruled}
\usepackage{tcolorbox}
\usepackage[dvipsnames,svgnames]{xcolor}
\usepackage{flushend}
\usepackage[numbers,sort&compress]{natbib}

\newcommand{\mylstset}{%
  \mylstsetCommon{}%
  \lstset{%
  }
}

\mylstset{}

\begin{document}


\title{
  \QoSFlow: Ensuring Service Quality of Distributed Workflows Using Interpretable Sensitivity Models
  }




\author{%
  \IEEEauthorblockN{Md Hasanur Rashid$^1$, Jesun Firoz$^2$, Nathan R. Tallent$^2$, Luanzheng Guo$^2$, Meng Tang$^3$, Dong Dai$^1$}
  \IEEEauthorblockA{$^1$University of Delaware, $^2$Pacific Northwest National Laboratory, $^3$Illinois Institute of Technology\\
  Email: \{mrashid,dai\}@udel.edu, \{jesun.firoz,tallent,lenny.guo\}@pnnl.gov, mtang11@hawk.iit.edu
  }
}




\thispagestyle{plain}
\pagestyle{plain}
\ifthenelse{\boolean{maketitleAfterAllMeta}}{}{\maketitle}


\begin{abstract}

With the increasing importance of distributed scientific workflows, 
there is a critical need to ensure Quality of Service (QoS) constraints, such as minimizing time or limiting execution to resource subsets.
However, the unpredictable nature of workflow behavior, even with similar configurations, makes it difficult to provide QoS guarantees.
For effective reasoning about QoS scheduling, we introduce \QoSFlow, a performance modeling method that partitions a workflow's execution configuration space into regions with similar behavior. Each region groups configurations with comparable execution times according to a given statistical sensitivity, enabling efficient QoS-driven scheduling through analytical reasoning rather than exhaustive testing. Evaluation on three diverse workflows shows that \QoSFlow's execution recommendations outperform the best-performing standard heuristic by 27.38\%. Empirical validation confirms that \QoSFlow's recommended configurations consistently match measured execution outcomes across different QoS constraints.

\end{abstract}










\ifthenelse{\boolean{maketitleAfterAllMeta}}{\maketitle}{}


\section{Introduction}
\label{sec:1_into}
Distributed scientific workflows have become a cornerstone of modern scientific discovery, orchestrating complex data processing pipelines across heterogeneous computing resources. They play a vital role in driving the control logic of emerging \textit{automated experiments} and \textit{digital twins}~\cite{nas:workflows:2022,Ferreira:2024:Computer-workflow-frontiers}.
As these workflows become more pervasive, ensuring \textit{Quality of Service (QoS)}, such as meeting strict deadlines or limiting the use of computational and storage resources, has emerged as a critical challenge in high-performance computing (HPC) environments~\cite{rashid2025adaptbf}. The growing reliance on data-driven analytics further amplifies this need~\cite{nas:workflows:2022,Ferreira:2024:Computer-workflow-frontiers}.
The service for any individual datum
often takes the form of a workflow composed of multiple stages, such as data partitioning, cleansing, analysis, recombination, and inference~\cite{Topcuoglu:2002:scheduling,li2022faasflow,Chowdhury:2022:IPDPS-dfman}.
This core pattern is reapplied across a dataset or series of experiments, possibly with optional components.
Depending on the use case, different QoS objectives may arise, such as achieving the shortest possible execution time, minimizing resource consumption, or maintaining high performance while operating within a restricted subset of resources.
However, workflow performance remains difficult to predict and control. Scientific workflows can exhibit large variability depending on scheduling strategies, task parallelism, and storage assignments~\cite{Topcuoglu:2002:scheduling,Ahn:2022:flux-runtime-feedback,Mayer:2017:tensorflow-critical-path,Wu:2023:path-metis,Buttazzo:partition-real-time,li2022faasflow,Chowdhury:2022:IPDPS-dfman,Herbein:2016:io-aware-scheduling,Fauzia:2013:tiling-data-locality,tang2016optimized,wang2014optimizing,Firoz+:2025:SSDBM-fastflow,Lee+:2025:IPDPS-flowforecaster}. Consequently, guaranteeing QoS in distributed scientific workflows remains an open and pressing research problem.

While prior works~\cite{yu2006scheduling,tolosana2011characterizing} have explored scientific workflow scheduling through heuristics (genetic algorithms) for deadline, budget constraints, and resilience, and recent performance analysis tools like LLAMP~\cite{shen2024llamp} and EDAN~\cite{shen2025edan} have provided deep insights into network latency and memory sensitivity for individual applications, significant gaps remain in workflow-level QoS management. In particular, \emph{existing optimization approaches focus on finding good configurations without explaining why they work or how sensitive they are to parameter changes}. Similarly, \emph{application-level analysis tools cannot capture the complex dataflow dependencies 
among workflow tasks.} While Krol et al.~\cite{krol2016workflow} developed workflow performance profiling with sensitivity analysis to characterize how input parameters affect resource consumption over time, their approach 
provides only retrospective sensitivity insights rather than actionable guidance for selecting optimal configurations to meet specific QoS objectives. There has been limited work on predicting scheduling configurations that can meet specific QoS constraints while providing interpretable guidance about which decisions are critical versus flexible. Previous QoS-aware workflow scheduling approaches~\cite{abrishami2013deadline,malawski2015algorithms} took a \emph{compute-centric} perspective that overlooks storage tier heterogeneity, I/O access patterns, and data flow characteristics, whereas our goal is to take an \emph{I/O-centric} perspective essential for modern data-intensive workflows.


In this work, we focus on \emph{I/O-intensive, data-driven scientific workflows}, where storage-tier selection and data movement can significantly influence end-to-end performance~\cite{rashid2025dial,rashid2023iopathtune,dong2025rl4sys,egersdoerfer2024understanding}. Different workflow stages can prefer different tiers such as remote parallel file systems, node-local SSDs, and memory-backed tmpFS, but crossing tiers introduce stage-in and stage-out costs that can shift bottlenecks across the workflow execution graph. This paper asks the following question: \emph{Is it possible to rapidly infer how a workflow's performance changes dependent on concurrency and storage-tier choices, and then use that knowledge to enforce explicit QoS constraints?}
This is challenging because workflow tasks can behave differently under different mappings of concurrency and storage tier, and the tasks in a workflow form an execution DAG (Directed Acyclic Graph)~\cite{deelman2015pegasus} whose critical path may \emph{shift unexpectedly between similar execution configurations}. Consequently, ensuring a workflow's QoS must address the following challenges:
(1) accurately predict which set of DAG paths will determine a workflow's end-to-end execution time (makespan) under different execution configurations;
(2) distinguish with statistical confidence among sets of configurations and DAG paths; 
(3) translate high-level QoS objectives into good feasible configurations; 
(4) and provide interpretable guidance about which configuration options are critical versus flexible.

This paper demonstrates how to form interpretable QoS models for workflows. We present \QoSFlow, an interpretable QoS modeling method that enables reasoning about a workflow's dataflow performance sensitivity under different scheduling configurations. 
When given a QoS objective and tolerance, the completed model predicts a superior execution configuration, including per-stage storage-tier assignment and concurrency settings, to meet the objective.
The system uses sensitivity analysis to partition the execution space into statistically distinct path-configuration groups that achieve similar performance under different constraints.
Each path represents data flow with analytic expressions that associate application input parameters (file size, access size) with analytic dataflow expressions.
The models are interpretable for three reasons:
(a) models represent performance as a \emph{known path} through the workflow's DAG;
(b) each path's dataflow is represented with \emph{analytic expressions};
(c) models quantify how much \emph{each configuration setting contributes} to performance.
Rather than relying on a myriad of workflow executions, the model is formed by reasoning about the execution space using emulation.
The emulation relies on only a few workflow executions, which is possible because of the analytic dataflow expressions.
Our approach builds on two complementary capabilities developed in prior work. First, we leverage workflow template construction and scale projection to instantiate a workflow's structure and demand at target scales~\cite{FlowForecasterRepo}. Second, we reuse storage-tier performance models based on a constrained set of profiles for individual workflow stages~\cite{Tang+:2025:DPM}. Building on these inputs, \QoSFlow composes stage-level estimates over the \emph{entire workflow DAG} and partitions the DAG-configuration space into interpretable regions, enabling constraint-driven configuration selection.

\COMPRESS{We evaluate \QoSFlow on three representative workflows (\texttt{1kgenome}~\cite{clarke20121000}, \texttt{PyFLEXTRKR}~\cite{feng2023pyflextrkr}, and \texttt{DeepDriveMD}~\cite{lee2019deepdrivemd}) and find that it partitions the configuration space into interpretable QoS regions, yielding a staircase-like ordering that simple storage heuristics fail to recover. Rendering region rules on the workflow DAG exposes which tasks are fixed to a storage resource and where flexibility remains. We observe \QoSFlow consistently produces tight, well-separated bands at all scales. A region-level decomposition of the critical path into resource utilization clarifies why regions differ and identifies critical assignments that influence end-to-end time. We further validate that \QoSFlow turns explicit QoS requests into actionable recommendations. Taken together, these observations demonstrate that \QoSFlow provides interpretable, scale-aware, and constraint-ready guidance for storage placement and concurrency choices.}

Our evaluation of \QoSFlow on three diverse workflows (\texttt{1kgenome}, \texttt{PyFLEXTRKR}, and \texttt{DeepDriveMD}) demonstrates that it successfully partitions the configuration space into interpretable QoS regions with distinctive staircase ordering that baseline greedy heuristics cannot achieve and outperforms the best performing one by 27.38\%. The approach consistently generates tight, well-separated performance bands across multiple scales, while region-level decomposition of critical paths shows the cost drivers behind performance differences and identifies storage assignments critical to 
workflow makespan. Finally, validation experiments confirm that \QoSFlow effectively translates explicit QoS constraints into actionable scheduling recommendations. These results establish \QoSFlow as an interpretable, scale-aware, and constraint-ready framework for workflow storage placement and concurrency optimization.

\myparagraph{Contributions}
%
We make the following contributions:
\begin{myitemize}

\item A QoS execution method inspired by sensitivity analysis that translates high-level QoS constraints into high-quality feasible execution configurations (schedule, task parallelism, storage tier assignment). 
\QoSFlow partitions a workflow's configuration space into performance regions using \emph{novel sensitivity analysis}.

\item \emph{Interpretable} models that 
distinguish between groups of execution configurations (regions);  highlight the key workflow paths that determine performance;
distinguish which configuration parameters are critical vs. flexible;
and in turn explain the critical path's performance using analytical dataflow expressions.

\item Validation demonstrating well-separated interpretable performance regions (better than the best performing baseline by 27.38\%), 
and experimental verification that \QoSFlow's predicted optimal configurations align with observed workflow performance across three representative workflows.
  
\end{myitemize}


\section{Overview}
\label{sec:2_overview}
\begin{figure}[t!]
  \centering
  \includegraphics[width=0.8\columnwidth]{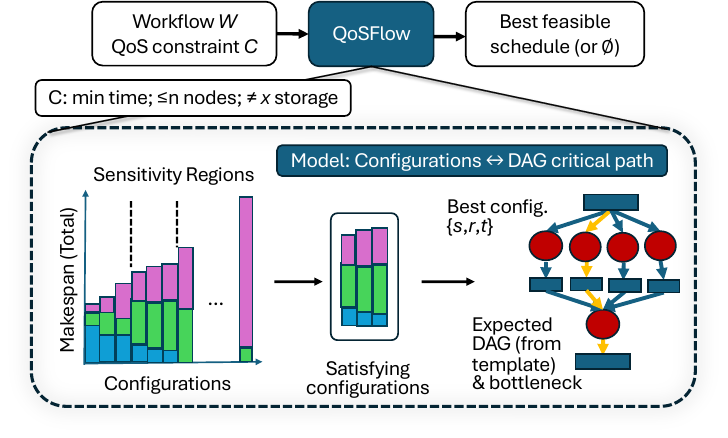}
  \caption{\QoSFlow system overview: Given a workflow and QoS constraints (e.g., minimize time with node/storage limits), \QoSFlow uses sensitivity analysis to map configurations to DAG critical paths and outputs the best feasible schedule that satisfies the constraints.
  }
  \label{fig:overview_flowqos}
\vspace{-1em}  
\end{figure}

Workflows consist of multiple applications that produce and consume data so as to form execution dependencies represented with directed acyclic graphs (DAGs). 
The DAG contains two types of vertices: \textit{data} and \textit{task} vertices.
Our objective is to build interpretable QoS models that identify which storage decisions critically impact execution time, partition the configuration space into performance regions, and translate high-level QoS objectives (time bounds, cost limits, performance targets) into concrete resource assignments. 

\myparagraph{Problem statement} Given a workflow $W$ and potential resources $R$, a scheduling problem typically finds a schedule (or scheduling policy) $S$ such that
\[S(W, R) = \max \text{\{schedule $s$, resources $r$, time $t$\}}\]
where “max” typically means sufficiently good rather than optimal; often the measure is $t$.
In contrast, our goal is a good QoS model $Q$ such that, given an execution constraint $C$ (e.g., time, resource usage) and a \emph{sensitivity threshold} $\varepsilon$, we find a \textbf{region of performance-equivalent configurations} rather than a single configuration. Configurations within a region achieve similar makespans (within $\varepsilon$ relative difference), enabling flexible scheduling decisions that distinguish critical choices from “don’t care” assignments. 
Let $y(x)$ denote the makespan of configuration $x=(s,r,t)$ and let $\widetilde y_i$ be the median makespan of region $R_i$.
We formalize this as:
\begin{align}
Q(W, C, \varepsilon) &= \arg\max_{\{s,r,t\}} S(W, R) \label{eq:qos_model} \\
\text{s.t. } \{s, r, t\} &\text{ satisfies } C,\quad\forall x, x' \in R_i: \frac{|\,y(x)-y(x')\,|}{\widetilde y_i} < \varepsilon \notag
\end{align}

\Cref{fig:overview_flowqos} illustrates the challenge.
Given a QoS constraint $C$ for $W$, we aim to understand the impact of resource selection on $W$'s performance through sensitivity analysis, decomposing each configuration’s contribution to performance and mapping configurations to DAG critical paths. This analysis enables us to identify satisfying configurations within the given constraints and to output a set of best feasible schedules (or $\emptyset$ if no feasible solution exists) that meet the QoS requirements.

\myparagraph{Challenges} Effective QoS-driven workflow scheduling through sensitivity analysis must address four key challenges:
\begin{myChallenge}
\item \label{challenge:scale} \textbf{Path-dependent behavior.} 
The critical path, the sequence of operations limiting total execution time (makespan), can shift entirely as resources (compute nodes, storage) change. Different storage configurations may activate different execution paths through the workflow DAG, altering which stages become bottlenecks. We need models that accurately predict this behavior (i.e. \textit{which critical path}) without requiring execution.

\item \label{challenge:sensitivity} \textbf{Resource sensitivity.} 
Not all configurations have equal impact on performance. 
\textit{How much impact} each one has may vary significantly depending on workload characteristics, data movement patterns, and resource utilization. 
Without systematic analysis, it's unclear which decisions require careful optimization versus which offer flexibility.

\item \label{challenge:qos} \textbf{Translating QoS conditions to schedules.} Users specify goals in terms of time bounds, cost limits, or tradeoff preferences. The system must translate high-level QoS requirements into specific, implementable configurations while providing guarantees about expected performance.

\item \label{challenge:interpret} \textbf{Interpretability of results.} Scheduling decisions must be explainable: why predicted configuration satisfies QoS as opposed to alternative configurations? What critical paths will dictate the performance? Which storage choices are critical, which are flexible, and how will performance change if conditions vary?
\end{myChallenge}

\myparagraph{Approach} \QoSFlow addresses these challenges through a sensitivity-driven modeling approach. \Cref{fig:overview_flowqos} shows our end-to-end system. The system takes as input a QoS constraint (time, resources) and workflow parameters $X$ (resource configurations, task scale, data scale), then outputs the best feasible schedule. The approach consists of three major phases:

\myparagraph{Phase 1: Workflow modeling and emulation} From a few empirical executions at different scales, we construct a DAG template with analytical scaling rules. This template projects the complete workflow DAG, including data volumes, access patterns, and dependencies, at any target scale without requiring actual execution (C1).

\myparagraph{Phase 2: Performance estimation and sensitivity analysis} We characterize available storage tiers by profiling their performance under various I/O conditions. For the projected DAG, we match each stage's I/O patterns to these storage profiles to estimate performance across all feasible configurations. We identify critical paths and perform sensitivity analysis to distinguish high-impact assignments from flexible ones (C2). 

\myparagraph{Phase 3: Region clustering and QoS-driven selection} We partition configurations into interpretable performance regions based on their sensitivity profiles and critical path characteristics. Given user QoS constraints, \QoSFlow recommends schedules that optimize high-sensitivity assignments on critical paths while exploiting flexibility elsewhere (C3, C4). Each region represents a set of configurations that achieve similar performance under specific conditions, enabling rapid QoS-driven configuration selection.
Importantly, this analysis scales to very large configuration spaces because of the focus on similar regions.

\section{Methodology}
\label{sec:3_methodology}

\begin{figure}[t]
  \centering
  \includegraphics[width=\columnwidth]{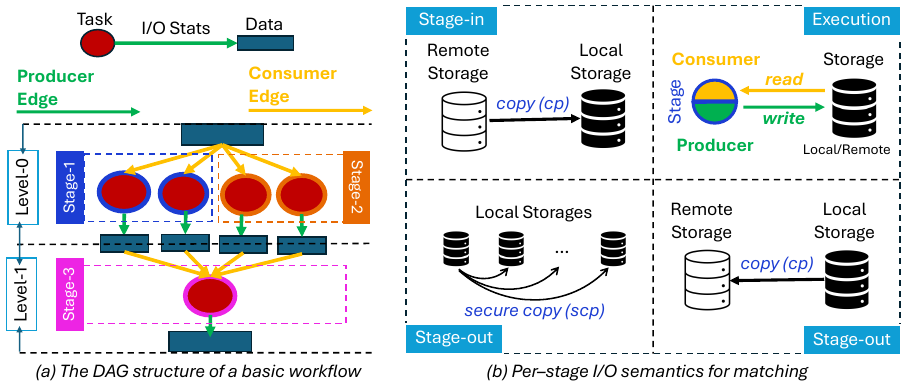}
  \caption{(a) A workflow DAG and (b) I/O semantics.}
  \label{fig:stage-io-semantics}
  \vspace{-1em} 
\end{figure}

\begin{figure*}[t]
  \centering
  \includegraphics[width=0.8\linewidth]{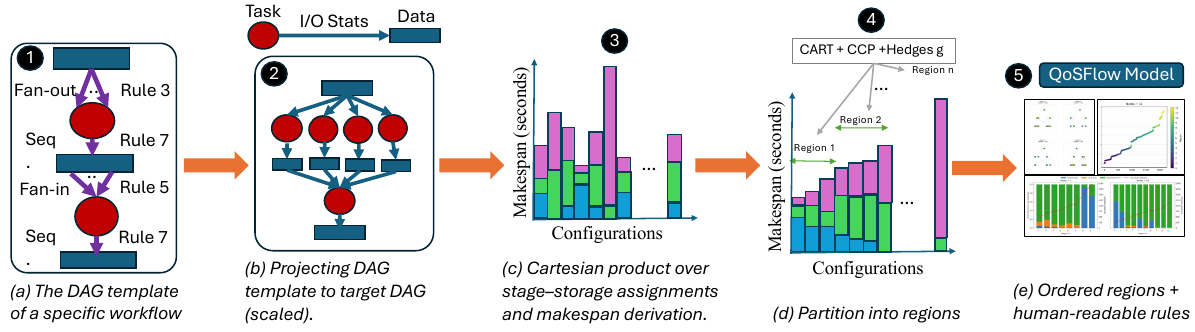}
  \caption{\QoSFlow methodology: (1) construct workflow DAG template, (2) project to target scale with I/O statistics, (3) enumerate all stage-storage configurations and compute critical-path makespans, (4) 
  partition configuration space into performance regions, and (5) generate interpretable QoS-driven scheduling rules.
  }
  \label{fig:overview}
\vspace{-1em}   
\end{figure*}

This section details \QoSFlow's \textit{sensitivity-driven} approach to building  \textit{interpretable workflow QoS models}.
The method revolves around a deep understanding of a workflow's DAG.
In a \textit{workflow DAG}, vertices (data and task vertices) and edges are annotated with property values representing execution and data flow statistics such as data access type, size, rates, reuse etc. The directed flow edges in the DAG represent producer-consumer relations. A \textit{producer} is an edge from a task to a data vertex; a \textit{consumer} is the reverse edge (i.e. from data to task). Workflows consist of multiple applications, henceforth referred to as \textit{stages}. Each stage is mapped to a \textit{level} of a workflow DAG. Figure~\ref{fig:stage-io-semantics}\textbf{(a)} illustrates this structure: tasks (circles) and data (rectangles) are grouped into stages aligned with levels, 
with producer edges (task$\to$data, green) and  consumer edges (data$\to$task, gold).
\Cref{fig:overview} illustrates the \QoSFlow method, which is discussed in the following subsections.
\COMPRESS{
\cref{subsec:preprocess-extraction} addresses the construction of the workflow-specific DAG template, applying this DAG template to project workflow structure at a target scale, and estimate per–edge I/O demand and stage performance under different storage configurations (\cref{fig:overview}, \circleb{1}-\circleb{2}). In \cref{subsec:enumerate-critical}, we enumerate all feasible stage–storage assignments and compute their critical-path makespans (\circleb{3}). 
We then perform sensitivity analysis to distinguish high-impact storage assignments from flexible ones in \cref{subsec:critical-analysis}. In \cref{subsec:region-clustering}, we partition the configuration space into interpretable regions using CART with cost–complexity pruning and variance-aware adjacent separation metrics (\circleb{4}). 
In \cref{subsec:qos-recommendation}, we translate these ordered regions into QoS-aware configuration recommendations and scheduling guidance (\circleb{5}).
}

\subsection{Model workflow characteristics}
\label{subsec:preprocess-extraction}
\myparagraph{Workflow DAG template construction}
As the first step (\circleb{1} in \cref{fig:overview}), we construct a representative \textit{DAG template} of a  workflow of interest using topological analysis that discovers the \textit{core graph}~\cite{Hell:1992:graph-core}, a minimal set of common and repeated \textit{subpatterns} that represent branches, joins, sequences, and loops in workflow DAGs.
This involves collecting a small set of workflow DAGs (3-5) with statistics (data volume, access size, accesses)  at different scales. 

To infer the scaling pattern from these statistics, we then employ an analytical \textit{rule-based} approach to explain how the subpatterns in the workflow scale and annotate each edge in the DAG template with a matching scaling rule. 
The rules describe how graph structure and edge properties change as inputs vary, such as how increasing the number of consumers processing a fixed dataset leads to proportional decreases in data volume along consumer edges. These rules are matched to edges of the workflow DAG template. The representative DAG template thus identifies common structural patterns and scaling behaviors. 
This DAG template serves as the fundamental building block for creating interpretable predictive models that can explain scaling trends across different workflow scales. To implement this step we leverage an  open-source tool~\cite{FlowForecasterRepo}.

\myparagraph{Projection of workflow demand at a target scale}
Once the DAG template, representing common substructures and their scaling patterns (rules) of a workflow, is constructed, it is used to project (instantiate) a workflow execution DAG at a target scale (either scaling tasks or data, or both). This is done by applying the scaling rules associated with each edge of the DAG template that dictates dataflow scaling. 
For example, a rule may dictate that as the input data is doubled while keeping the number of tasks fixed, the data volume transferred along each edge of the graph at the target scale also doubles while the access size remains fixed. 
In this way, \textit{entire DAG substructures at a target scale with annotated edge properties (resource demand) can be predicted without requiring actual workflow execution}, enabling pre-execution reasoning about performance bottlenecks and resource allocation (\circleb{2} in \cref{fig:overview}).

\noindent
\textbf{Dataflow performance projection.}
After estimating the  \textit{dataflow demand} at  target scale using the DAG template and rules, a \textit{dataflow performance matching} tool~\cite{Tang+:2025:DPM} projects 
workflow stages I/O performance 
under various dataflow resource configurations. 
This projection involves 
2 steps.

First, performance of various storage tier is characterized. 
For this, IOR~\cite{ior_tool} benchmarks with carefully selected I/O building blocks are utilized to simulate various I/O operations that storage systems encounter under different workloads, generating detailed performance profiles for each storage tier. These \textit{storage tier profiles} capture combinations of key metrics like: 
data access patterns (sequential/random, read/write), average I/O transfer sizes per operation, 
task parallelism,
and total I/O size. Importantly, this system-wide storage tier characterization is done only once, independent of any specific workflow. 

Second, the  \textit{matching} method combines the pre-measured storage tier profiles and instantiated workflow DAG at the target scale, 
calculating a performance score for each stage within the workflow DAG. Each score represents the weighted estimated I/O times for different storage-parallelism configurations, where lower scores indicate more efficient scheduling decisions with lower predicted I/O overhead.

Figure~\ref{fig:stage-io-semantics}\textbf{(b)} illustrates how
we decompose I/O into three components to accurately model data movement costs during workflow execution: (i) \textbf{stage-in} handles pre-execution data relocation from remote storage, (ii) \textbf{execution} captures actual task I/O operations where consumers read and producers write to assigned storage while enforcing data locality, and (iii) \textbf{stage-out} manages post-execution data redistribution across storage tiers. Performance estimates for these components are derived from storage profiles matching the producer-consumer edge's access patterns, transfer characteristics, and concurrency levels, then aggregated across all workflow stages to compute total makespan scores for complete configurations.

\subsection{Critical Path Identification and Sensitivity Analysis}
\label{subsec:enumerate-critical}

Having established individual stage-storage performance scores, the system systematically enumerates the global configuration space by generating all possible stage-to-storage assignments across the entire workflow DAG (\cref{fig:overview}, \circleb{3}). This exhaustive approach creates a Cartesian product of all stage-storage tier combinations.
For each configuration, the system evaluates the DAG level-by-level in topological order. Within each level, multiple stages may execute in parallel, but the level's completion time is determined by the slowest stage (straggler). The system computes the maximum of the three I/O component times (stage-in, execution, stage-out) within each level, then sums these per-level maxima across all DAG levels to obtain the total makespan.
During this evaluation process, the system tracks \textit{critical components}—the specific stage-storage pairs that determine each level's maximum execution time. These bottleneck operations form the critical path trace, identifying which storage assignments have the greatest impact on overall workflow execution time.

The critical path identification reveals that not all storage assignments equally impact workflow performance. Understanding the \textit{degree} of such impact requires careful consideration. Additionally, multiple storage configurations may yield comparable execution times for the same critical path, providing flexibility in resource allocation decisions.
To this end, we employ \textit{sensitivity analysis} ~\cite{da2021basics,smith2024uncertainty,saltelli2004sensitivity,borgonovo2016sensitivity} -- a systematic approach to quantify how variations in input parameters of a workflow (storage assignments, parallelism levels, data sizes) affect output metrics (execution time, resource cost). 
This enables us to distinguish among configuration choices that significantly influence performance vs. those with minimal impact.
\COMPRESS{, thereby prioritizing optimization efforts on the most consequential decisions while allowing flexibility in less critical assignments.}

\COMPRESS{
Mathematically, sensitivity analysis can be framed as computing the partial derivatives of objective functions with respect to decision variables.
}
\COMPRESS{
\textit{Global sensitivity analysis}~\cite{borgonovo2008sensitivity,borgonovo2016sensitivity} examines how variations across the entire input space affect outputs, revealing which parameters fundamentally drive system behavior and identifying non-linear interactions between multiple factors. In contrast, \textit{local sensitivity analysis}~\cite{pianosi2016sensitivity} evaluates the impact of small perturbations around a specific configuration, quantifying how robust a particular solution is to minor changes in conditions. 
}
For workflow scheduling, global sensitivity analysis~\cite{borgonovo2008sensitivity,borgonovo2016sensitivity} helps identify which 
stages are inherently critical
to overall performance across different scales and input characteristics regardless of specific storage tier assignments. In contrast, local sensitivity analysis~\cite{pianosi2016sensitivity} assesses the impact of small perturbations around a specific promising configuration, quantifying how robust a particular solution is to minor changes in conditions -- for instance, whether a critical path remains critical when storage performance varies slightly, or whether small changes in data size cause critical path transitions between different execution sequences.

Applying this framework to the enumerated configurations, we perform three complementary analyses built upon the critical path insights. \emph{First}, sensitivity classification categorizes storage assignments based on their impact on overall execution time: assignments on critical paths typically show high sensitivity, while those on non-critical paths whose variation causes changes below a predefined threshold 
are marked as ``don't care'' decisions, offering flexibility for storage tier allocation without performance penalties. 
\emph{Second}, we identify 
\COMPRESS{ Pareto-optimal 
solutions 
that represent non-dominated trade-offs between execution time and resource cost, filtering out configurations that are strictly worse on both dimensions. This analysis often reveals}
 multiple storage configurations that yield comparable execution times -- a finding that reflects flat regions in the sensitivity  \textit{response surface}~\cite{borgonovo2016sensitivity} landscape where several assignments result in similar performance. \emph{Third}, robustness assessment evaluates how each highly-ranked configuration responds to variations in input data size, task parallelism, and storage performance characteristics, specifically examining whether critical path transitions occur and whether performance remains effective under realistic operational uncertainties. Together, these analyses produce a ranked set of configurations with explicit indicators of decision criticality and solution robustness informed by both critical path structure and sensitivity characteristics.

\subsection{Region-based Configuration Clustering}
\label{subsec:region-clustering}

The sensitivity analysis reveals important patterns in configuration space: certain storage assignments on critical paths strongly influence performance, while others in low-sensitivity regions offer flexibility; multiple configurations often achieve comparable execution times, suggesting the existence of \textit{performance-equivalent regions}.
To make these patterns explicit and actionable for scheduling decisions, \QoSFlow employs Classification and Regression Trees (CART)~\cite{breiman2017classification} to partition configuration space into interpretable performance regions (\cref{fig:overview}, \circleb{4}).

CART is a decision tree algorithm that recursively splits the data based on feature values to predict a target variable. 
It builds regression trees for numeric targets (like performance outcomes in our case) and classification trees for categorical targets. During training, CART finds optimal splits that minimize prediction error and continues recursively until stopping criteria are met (e.g., maximum depth or minimum samples per leaf). Each terminal leaf represents a performance region, but without careful stopping criteria, overfitting risks creating too many tiny regions. Cost-complexity pruning (CCP)~\cite{breiman2017classification} addresses this by defining parameter $\alpha$ that penalizes tree size -- smaller $\alpha$ creates deeper trees while larger $\alpha$ creates shallower trees (complexity vs. generalization trade-off). The system requires a systematic approach to search for the optimal $\alpha$ that maximizes the objective function for meaningful region identification.

\begin{figure}[t]
  \centering
  \includegraphics[width=0.8\columnwidth]{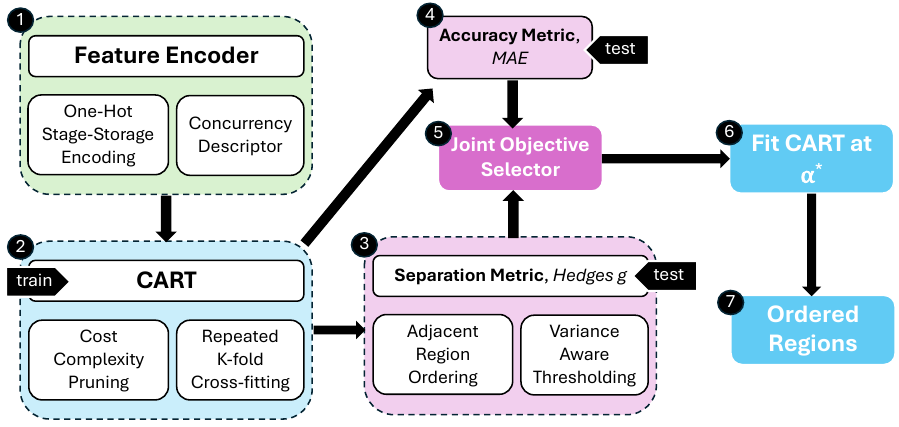}
  \caption{\QoSFlow identification of performance regions with CART}
  \label{fig:cart}
\vspace{-1em}  
\end{figure}

\myparagraph{Overview of the pipeline} \Cref{fig:cart} illustrates our CART-based region identification pipeline.
First, we encode the configuration into features to allow split on features while training CART \circleb{1}. A  feature encoder performs one-hot encoding for per–stage storage tier choices (categorical) and retains the scale (numerical) as it is (if present). We include the scale to enable scale-aware split for CART.
Next, we train CART to predict region makespan with cost–complexity pruning and repeated $K$-fold cross-fitting \circleb{2}. The cross-fitting with explicit train/test splits helps avoid data leakage and the repetition creates stability.
On each \textit{test} fold, we compute an adjacent-region separation metric based on Hedges' $g$~\cite{hedges1982estimation} with variance-aware thresholding \circleb{3}.
We then compute CART's prediction error on regions' makespan using mean absolute error (MAE) on each \textit{test} fold \circleb{4}.
A joint objective combines the normalized separation and accuracy to choose the pruning level $\alpha^\star$ along the cost–complexity path \circleb{5}.
This helps us find a pruning level where the tree is both a good predictor and produces statistically meaningful, well-separated regions. We later discuss in detail how we calculate the metrics and derive the objective.
After finding the best pruning level ($\alpha^\star$), we refit CART on all data at selected $\alpha^\star$ \circleb{6}.
We then order final regions and their configurations on median makespan of each region to output \emph{ordered regions} \circleb{7}.

\myparagraph{Balancing Predictive Accuracy and Region Interpretability} We now formalize the region identification process and the $\alpha$ selection criterion. Our region selection strategy jointly optimizes two critical criteria: predictive accuracy and region separability. For predictive accuracy computation, we measure prediction quality using MAE on makespan predictions. For each  cost--complexity penalty $\alpha$, we aggregate MAE scores across all cross-validation folds using the median to ensure robustness against timing noise, yielding $\mathrm{mae}_{\text{med}}(\alpha)$. For adjacent-region separation metric, to quantify region quality, we develop a variance-aware separation metric based on Hedges' g effect size. For configurations on a held-out set, let $Y_i$ denote the makespan observations in leaf $i$, ordered by median performance. For each adjacent pair $(i,j)$, we compute:

\begin{align}
\nu_{ij} &= n_i+n_j-2, & J(\nu) &= 1-\frac{3}{4\nu-1},\\
g_{ij} &= J(\nu_{ij})\,\frac{\lvert \bar y_i-\bar y_j\rvert}{\sqrt{\tfrac{1}{2}(s_i^2+s_j^2)}}. \label{eq:hedges_g}
\end{align}

Where $n_i=|Y_i|$, $\bar y_i$ the mean of $Y_i$,  $s_i$ its standard deviation (analogously for $j$) and $g_{ij}$ denotes \textit{effect size} with Hedges’ small\mbox{-}sample correction. This effect size captures the standardized difference between adjacent regions while correcting for small sample bias. 

Next, to distinguish meaningful separations from noise-driven gaps, we implement adaptive thresholding based on pooled coefficient of variation. In particular,  let $\mathrm{CV}_i=s_i/\bar y_i$ be the leaf\mbox{-}level coefficient of variation, and define a pooled variability index
\begin{equation}
\mathrm{CV}_{\text{pooled}}=\sqrt{\tfrac{1}{2}\bigl(\mathrm{CV}_i^2+\mathrm{CV}_j^2\bigr)}. \label{eq:pooled_cv}
\end{equation}
We then use $\mathrm{CV}_{\text{pooled}}$ to set a variance\mbox{-}adaptive threshold:
\begin{equation}
g_{\text{thr}}=\max\!\bigl(g_{\mathrm{floor}},\,\min(g_{\mathrm{cap}},\,\delta/\mathrm{CV}_{\text{pooled}})\bigr)
\end{equation}
where $g_{\mathrm{floor}}$ avoids declaring tiny gaps as ``separated'', $g_{\mathrm{cap}}$ prevents overly strict thresholds for stable regions, and $\delta$ encodes the minimal practically relevant relative difference.

We compute the final separation score by weighting each adjacent pair by the harmonic mean of sample sizes and aggregating over pairs that exceed the threshold:
\begin{equation}
\operatorname{sep}(\alpha)=
\frac{\sum_{(i,j)\ \text{adjacent}:\ g_{ij}\ge g_{\text{thr}}} g_{ij}\,w_{ij}}
{\sum_{(i,j)\ \text{adjacent}} w_{ij}}. \label{eq:sep_metric}
\end{equation}
This weighting down-weights pairs with very small sample sizes while emphasizing well-supported separations.

\textit{Optimal pruning selection.}
The optimal pruning level $\alpha^\star$ emerges from a joint objective that balances accuracy and separability. We normalize both $\mathrm{mae}_{\text{med}}(\alpha)$ and $\mathrm{sep}_{\text{med}}(\alpha)$ across all candidate $\alpha$ values  to obtain their normalized values,  $\widetilde{\mathrm{sep}}(\alpha)$ and $\widetilde{\mathrm{mae}}(\alpha)$.
The joint objective combines these normalized metrics with equal weighting:
\begin{equation}
J(\alpha)=
w\,\widetilde{\mathrm{sep}}(\alpha)
+(1-w)\,\bigl(1-\widetilde{\mathrm{mae}}(\alpha)\bigr),
\quad w=0.5,
\label{eq:selection_objective}
\end{equation}
and choose $\alpha^\star\in\arg\max_\alpha J(\alpha)$. This avoids trees that predict well but yield trivial regions (low separation), as well as trees that over\mbox{-}segment on noisy gaps (poor error). By setting $w$ to $0.5$, we place equal weight 
to both metrics to obtain $J(\alpha)$.


\textit{Final regions and ordering.}
Using the optimal $\alpha^\star$  we refit CART to generate  the final set of regions 
(\circleb{4} in \cref{fig:overview}). 
Regions and their configurations are ordered by their \emph{median} makespan for robustness. The result is a staircase-like policy map: low within-region variance, and clear between-region performance steps.
(\circleb{5} in \cref{fig:overview}).


\textit{Complexity.}
Let $N$ be the number of evaluated configurations, $p$ the number of input features (one-hot stage--storage indicators plus any concurrency descriptors), $D$ the learned tree depth, $K$ the number of outer folds, $R$ the number of repeated cross-validation runs, and $m$ the number of leaves (regions) induced on a held-out split. Let $A$ denote the number of candidate cost--complexity pruning parameters $\alpha$ on the pruning path.
For a fixed $(\alpha,\text{fold})$, fitting a CART regressor costs $O(N\,p\log N)$ time and $O(N\,p)$ memory in practice, and evaluating the held-out fold costs $O(N\,D)$ for predictions. The separation metric assigns held-out configurations to leaves and aggregates per-leaf statistics in $O(N)$, then orders the $m$ leaves by their held-out medians in $O(m\log m)$ and evaluates the $(m-1)$ adjacent pairs in $O(m)$.
Therefore, repeated $K$-fold selection over the $A$ candidate pruned subtrees costs
\[
O\!\Big(R\,K\,A\,\big(N\,p\log N \;+\; N\,D \;+\; N \;+\; m\log m\big)\Big),
\]
and since $m\!\ll\!N$ typically and the learned trees are shallow, the dominant term is
\[
O\!\Big(R\,K\,A\,N\,p\log N\Big).
\]
The pruning path per split is linear in the size of the initially grown tree, so $A$ is at most the initial number of leaves on that split; in our implementation, selecting $\alpha^\star$ is a single sweep over this path by maximizing the joint objective $J(\alpha)$. After selection, refitting the final tree at $\alpha^\star$ is one additional CART fit on all data.
\emph{Crucially, region identification cost depends on the number of evaluated configurations ($N$) and feature dimensionality ($p$), and is independent of execution scale such as the number of allocated compute nodes, because configurations are defined by stage-to-storage and concurrency descriptors rather than node-level schedules.}

\textit{Downstream cost after region identification.}
Once $\alpha^\star$ is selected and the final pruned tree is refit, the model summarizes the $N$ evaluated configurations into $m$ interpretable regions (leaves), typically with $m\!\ll\!N$.
Downstream tasks operate on this compact representation: assigning a new configuration to a region requires a single tree traversal, which is $O(D)$ time, and producing region-level summaries or ordered region lists scales with the number of regions, for example $O(m\log m)$ to order regions by median makespan and $O(m)$ to scan adjacent region pairs.
As a result, repeated planning, interpretation, and reporting are governed by the number of regions and tree depth rather than by the number of evaluated configurations.

\subsection{QoS-driven Configuration Recommendation}
\label{subsec:qos-recommendation}
The CART-based partitioning and sensitivity analysis provide a comprehensive understanding of the configuration space, but practical scheduling requires translating this knowledge into actionable decisions that satisfy user-specified QoS requirements. This final component of \QoSFlow bridges the gap between analytical insights and operational constraints by mapping user requirements to appropriate storage configurations identified through the previous analysis stages (\cref{fig:overview}, \circleb{5}).

Users specify QoS requirements that may include maximum acceptable execution time, budget constraints, or preferred cost-performance trade-offs (such as prioritizing fastest execution vs. most cost-effective solution). These requirements are mapped to corresponding regions in CART-generated decision tree
to 
identify feasible configuration spaces. The tree traversal process is informed by the sensitivity analysis: high-sensitivity parameters identified earlier guide the primary decision branches, ensuring critical storage assignments on critical paths are selected to meet performance targets, while low-sensitivity ``don't care'' decisions in non-critical regions are resolved based on secondary criteria like cost or storage tier availability. 

Within the selected feasible region, the system recommends specific storage assignments that satisfy the QoS constraints while optimizing for the user's stated priorities. For users prioritizing execution time minimization, the recommendation selects configurations 
that achieve the fastest performance within budget constraints. For cost-conscious users, the system leverages the performance-equivalent regions identified during sensitivity classification to select the least expensive configuration that still meets time requirements, exploiting the flexibility in low-sensitivity assignments. 

\textit{Scalability of \QoSFlow.}
\QoSFlow achieves scalability by dramatically reducing the exploration space in three steps: (i) instead of testing every possible way to assign storage tiers to workflow stages (which grows exponentially), it uses sensitivity-driven region partitioning to collapse thousands of similar configurations into a small number of representative \textit{regions}; (ii) this region-building step is followed by constraint-based \textit{pruning} that filters out configurations that can't meet user requirements, and (iii) then \textit{analytical modeling} through DAG templates that predicts performance at larger scales without requiring actual execution. As resources and available configurations grow, this region-prune-predict approach provides more opportunities to satisfy user-specified thresholds while keeping the search space manageable.

\section{Evaluation}
\label{sec:4_evaluation}
We evaluate \QoSFlow on three representative scientific workflows to demonstrate its effectiveness in storage configuration optimization tailored to a QoS request. Our evaluation shows that \QoSFlow outperforms different heuristic policies in identifying and ordering performance regions. We then illustrate how regions output human-readable rules aligned with the workflow DAG semantics. For each workflow, we examine how \QoSFlow forms the QoS regions at different node scales and analyze how storage tier assignments influence total makespan within these regions. Finally, we empirically validate \QoSFlow's effectiveness in handling diverse QoS requests.

\myparagraph{Platform}
We conducted all experiments on 
an HPC cluster with the following configuration. 
Each compute node features \emph{dual AMD EPYC~7502} CPUs (64 cores total) with 256\,GB of eight-channel DDR4-3200 memory, 6 NVIDIA A100 GPUs with 80GB memory, and a \emph{512\,GB} node-local NVMe SSD, providing greater than 1 GB/s read/write performance. 
The compute nodes are interconnected with \emph{HDR-100 Mellanox InfiniBand} (100\,Gb/s). 
The platform provides multiple storage options including NFS, BeeGFS~\cite{herold2014introduction} with caching, and Ramdisk. For our evaluation, we choose one out of three possible storage tiers for workflow stage-storage assignment: BeeGFS (remote), node-local SSD (local), and a tmpFS based on main memory (local).

\begin{figure}[t!]
  \centering
  \includegraphics[width=0.8\columnwidth]{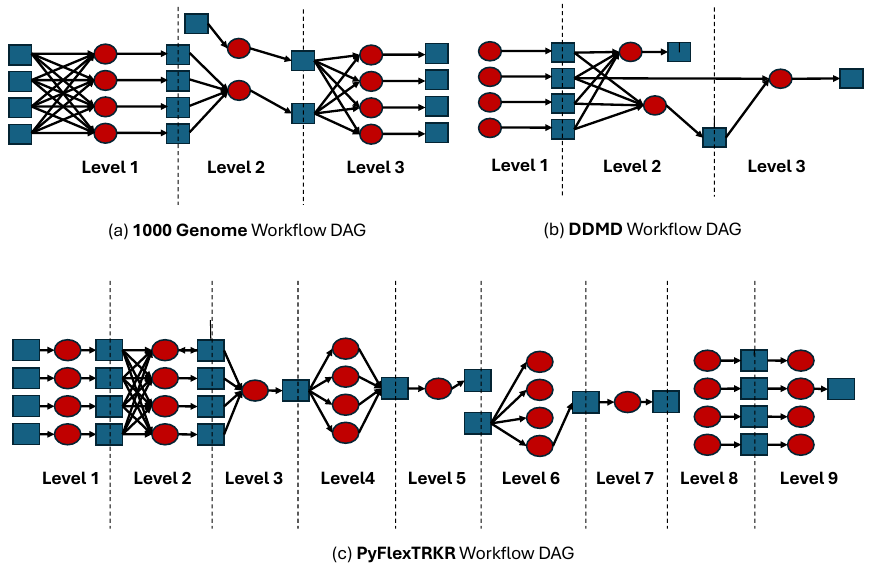}
  \caption{Representative workflow DAGs with annotated levels: (a) 1000 Genome, (b) DDMD, and (c) PyFLEXTRKR.}
  \label{fig:three-dags}
\vspace{-1em}  
\end{figure}

\subsection{Case Study: 1000 Genomes (1kgenome)}
\label{sec:wf-1kgenome}
The 1000 Genomes workflow~\cite{clarke20121000,1kG_git} is a data-intensive bioinformatics workflow that processes genomic data through five application stages: \texttt{individuals} (per–chromosome extraction and processing), \texttt{individuals\_merge} (aggregation across chromosomes), \texttt{sifting} (SNP SIFT scoring), \texttt{frequency} (overlap frequency computation), and \texttt{mutation\_overlap} (overlap of mutations). The workflow's  DAG  structure is illustrated in Fig.~\ref{fig:three-dags}a.



\myparagraphA{Policy–based orderings comparisons}
\label{sec:eval-policies}
HPC scheduling policies typically rely on established heuristics~\cite{pan2018h,zaharia2010delay,topcuoglu2002performance} for resource allocation decisions. To demonstrate \QoSFlow's effectiveness, we first compare our policy-induced orderings against representative baseline policies
of the \emph{same} configuration set with a fixed number of computing nodes (10) running \texttt{1kgenome} workflow. The configuration set consists of all possible combinations of the three storage-tier assignments across workflow stages. We determine the makespan of each configuration following the procedure described in \S~\ref{subsec:enumerate-critical}.
By maintaining a fixed node count and using the same  configuration set, we ensure that any structural patterns observed in the results reflect the policy's effectiveness rather than variations in the underlying data.
We evaluate four distinct policies, including \QoSFlow, for deriving ordered configurations:

\begin{itemize}[leftmargin=*]
    \item \textit{Fastest-Storage First (FSF).} 
    This policy prioritizes configurations that maximize the use of high-bandwidth storage tiers~\cite{pan2018h}. For each configuration, we count the number of stages assigned to the fastest storage tier and the number assigned to the second-fastest tier. Configurations are then ordered using descending lexicographic ordering, first maximizing fastest-tier usage, then second-fastest tier usage. This approach promotes broad deployment of fast storage while remaining agnostic to specific stage characteristics.
    \item \textit{Low-Transition Layout (LTL).} 
    This policy minimizes data movement between storage tiers by computing a transition score for each configuration~\cite{zaharia2010delay}. The score increments by one for each stage-boundary action that induces data movement (stage-in and stage-out operations as defined in \S~\ref{subsec:preprocess-extraction}. Configurations are ordered by ascending transition score, where lower scores indicate reduced inter-tier data movement and potentially better performance.
    \item \textit{Hybrid Heuristic (FSF $\oplus$ LTL).} 
    This policy combines the objectives of both FSF and LTL, balancing fast storage utilization against data movement penalties~\cite{topcuoglu2002performance}. Configurations rank higher when they deploy fast storage across more stages while simultaneously minimizing stage-in/out actions. The combined score is computed by rewarding fast media usage and penalizing boundary transitions, with configurations ordered by descending combined score.
    \item \textit{\QoSFlow.} 
    This approach employs our trained CART regression model to order configurations based on predicted performance and separability, as detailed in \S~\ref{subsec:region-clustering}.
\end{itemize}

\begin{figure}[t!]
  \centering
  \includegraphics[width=0.8\columnwidth]{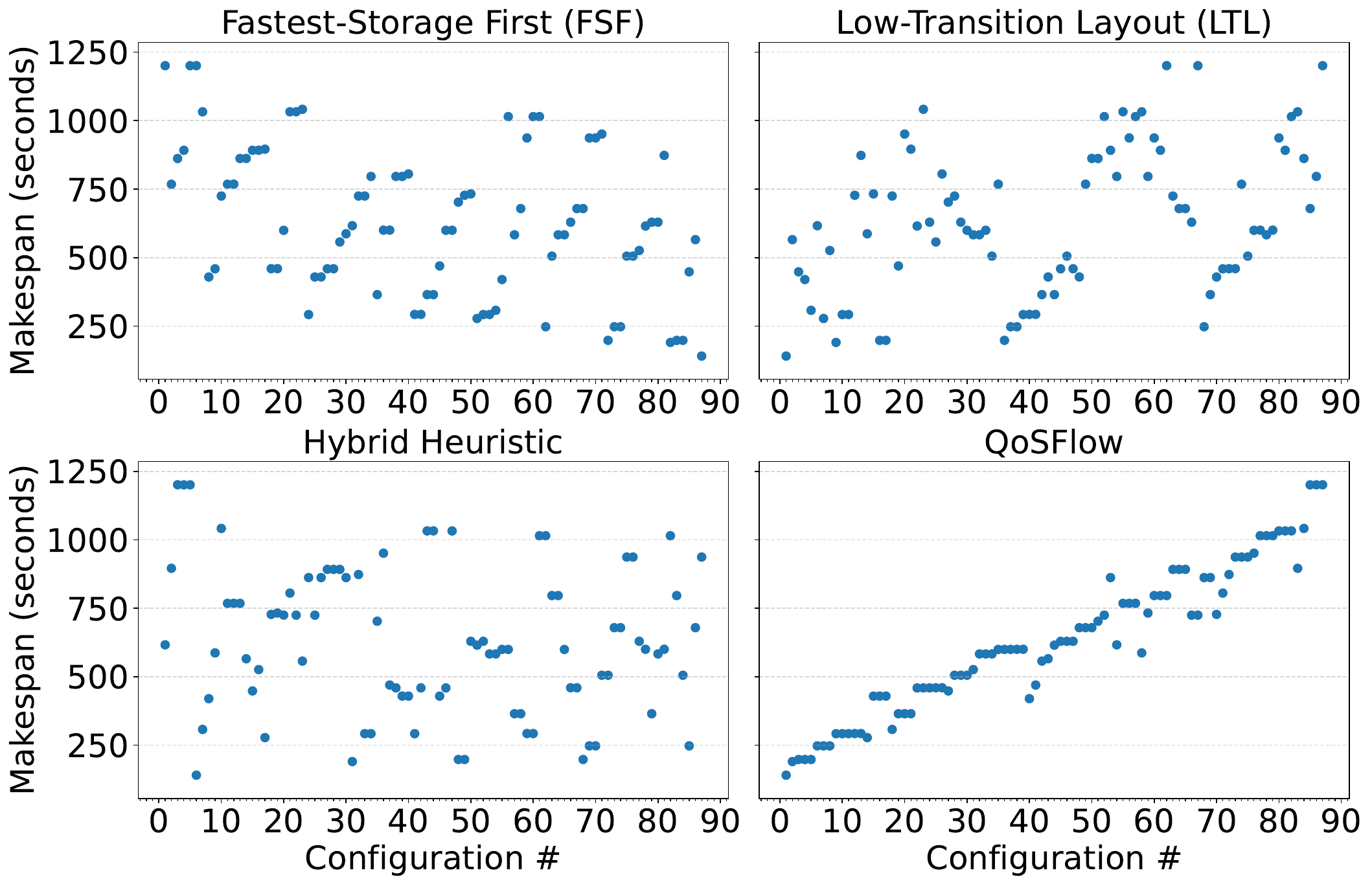}
  \caption{Configuration ranking quality comparison for \texttt{1kgenome} showing \QoSFlow's ability to create well-separated performance groups versus ineffective baseline heuristics.
  (10 compute  nodes).}
  \label{fig:policy-six}
  \vspace{-1em}
\end{figure}

\begin{table}[t]
  \centering
  \caption{\small Pairwise Concordance (PC) on \texttt{1kgenome} (10 nodes) and how much \QoSFlow is better than each baseline (in \%).}
  \label{tab:policy-pc}
  \small
  \setlength{\tabcolsep}{10pt}
  \begin{tabular}{lcc}
    \hline
    \textbf{Policy} & \textbf{PC} & \textbf{\% better by \QoSFlow} \\
    \hline
    FSF     & 0.378 & 153.04\% \\
    LTL     & 0.751 & \ \,27.38\% \\
    Hybrid  & 0.503 & \ \,90.12\% \\
    \QoSFlow & \textbf{0.956} & - \\
    \hline
  \end{tabular}
\vspace{-1em}  
\end{table}

\Cref{fig:policy-six} presents the ordering outcomes for configurations under each policy. The X-axis represents the  configuration index (with lower indices indicating better configurations according to each policy's criteria), while the Y-axis shows the actual workflow makespan for each configuration. 
From \cref{fig:policy-six}, we observe that, FSF, LTL, and hybrid policies produce configurations that are randomly scattered across the makespan range, indicating that the configurations in their ranking criteria 
that were deemed ``better'' by these policies fail to yield better makespan (lower is better). In contrast, the \QoSFlow ordering generates a distinctive staircase profile 
whose median values increase steadily from left to right.
We quantify ordering fidelity using \emph{Pairwise Concordance (PC)}~\cite{kendall1938new}, the fraction of configuration pairs that a policy orders in the same direction as measured makespan (1.0 is perfect; 0.5 is random; < 0.5 indicates anti-correlation). As shown in~\Cref{tab:policy-pc}, \QoSFlow achieves near-perfect concordance (PC\,=\,0.956) and is 27.4\% better than the best-performing heuristic (LTL).
Hence, \QoSFlow successfully orders configurations in a manner that closely approximates the ideal ranking obtained by directly sorting configurations by makespan, while maintaining the interpretability advantages of region-based grouping.

\begin{figure}[t!]
  \centering
  \includegraphics[width=0.6\columnwidth]{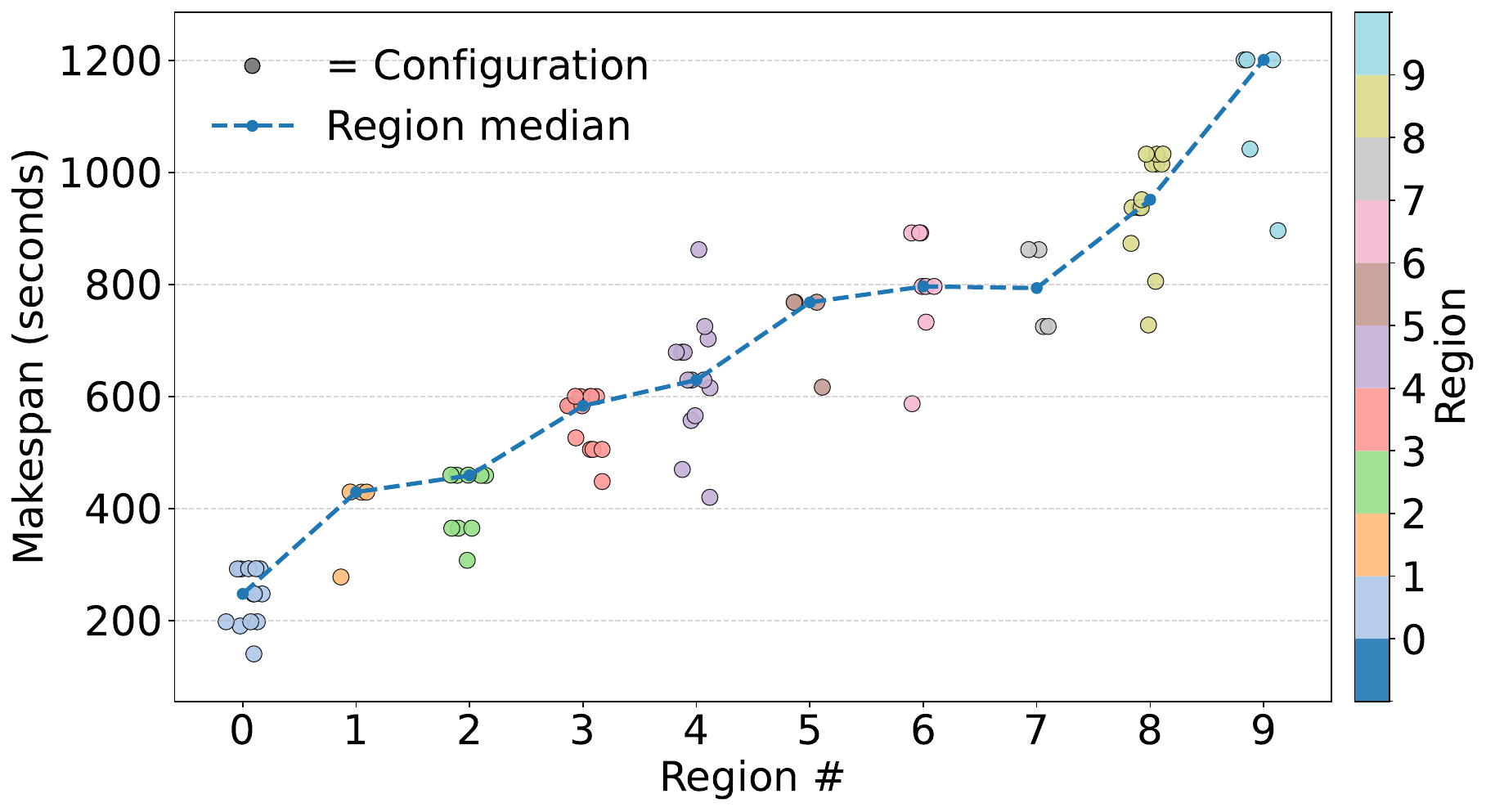}
  \caption{\QoSFlow-suggested QoS regions for \texttt{1kgenome} with 10 nodes.}
  \label{fig:region-clusters}
 \vspace{-1em} 
\end{figure}



\Cref{fig:region-clusters} illustrates how the configurations are clustered for each region in terms of makespan for \QoSFlow in \cref{fig:policy-six}. The regions are indexed by \QoSFlow according  to their median makespan values with lower indices corresponding to better-performing regions. The visualization confirms that \QoSFlow discovers performance clusters with tight intra-region dispersion and well-separated inter-region medians. This produces the staircase pattern across regions, a characteristic that the baseline heuristic orderings consistently fail to achieve (~\cref{fig:policy-six}).


\begin{tcolorbox}[colback=bg,left=1ex,top=1ex,boxsep=0ex,bottom=1ex,width=\linewidth]
\obs{} \QoSFlow-derived regions form clearly separated, low-variance performance groups that baseline heuristics cannot reliably identify or recover.
\end{tcolorbox}

\begin{figure*}[t!]
  \centering
  \includegraphics[width=0.8\textwidth]{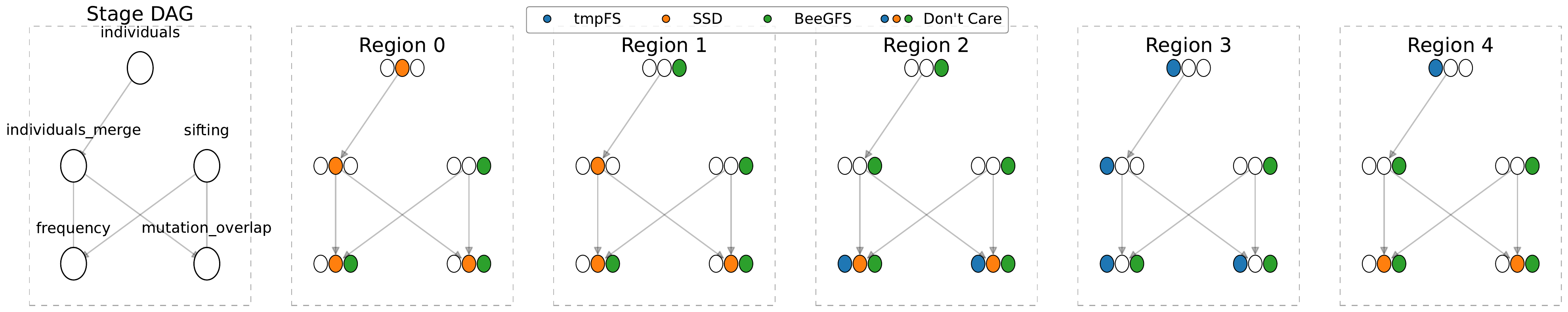}
  \caption{Interpretable region rules: QoS regions rendered over the workflow DAG with stage dependencies. Each workflow stage vertex is represented by a 3-position storage-set glyph (\textcolor{blue}{tmpFS}, \textcolor{orange}{SSD}, \textcolor{DarkGreen}{BeeGFS}), where filled markers indicate permitted storage tiers -- multiple filled markers show flexible assignments while single filled markers reveal fixed constraints across performance regions, ordered from best-performing (leftmost) to worst-performing configurations.
  }
  \label{fig:dag-panels}
 \vspace*{-1em} 
\end{figure*}

\myparagraphA{Interpretable QoS Regions}
\label{sec:eval-dag-glyphs}
\COMPRESS{While ranking regions by performance provides essential guidance, it alone is insufficient for practical scheduling and QoS management.} 
Users may also benefit from gaining
insights into what each region represents operationally: specifically, which workflow stages are constrained to particular storage tiers and where flexibility across tiers remains available. This semantic understanding enables users to make informed decisions among configurations that exhibit similar performance characteristics, allowing them to balance performance requirements with operational constraints and resource availability.


To address this need, we combine the quantitative performance separation view with a semantic rendering of \emph{region rules} directly on the workflow DAG. Each region rule defines the set of admissible storage tiers for every workflow stage: singleton sets indicate fixed storage assignments, while multi-element sets represent flexible tier choices that maintain equivalent performance within the region.

\Cref{fig:dag-panels} visualizes these rules using a compact notation where each workflow stage is represented by a three-position \emph{set-valued node glyph} corresponding to the storage tiers (tmpFS, SSD, BeeGFS from left to right). Within each glyph, filled markers indicate permitted storage tiers for that region, while empty markers denote excluded tiers. When all three positions are filled, the stage has complete flexibility across all storage options. The panels are ordered by increasing region median makespan, enabling direct comparison of storage constraints across performance levels.




The consistent DAG topology and node positioning across panels facilitates visual analysis of how storage constraints evolve with performance requirements. Users can immediately identify where storage assignments become fixed (singleton constraints) versus where flexibility remains available (multiple filled markers). For visual clarity, we present the top five regions in median makespan in \cref{fig:dag-panels}.


As a concrete illustration, consider the \texttt{1kgenome} workflow at 10 nodes shown in~\Cref{fig:dag-panels}. The leftmost region panel representing the best-performing configurations (lowest median execution time) demonstrates that \texttt{frequency} and \texttt{mutation} stages retain storage flexibility (two marks filled), allowing selection among SSD or BeeGFS.  Conversely, \texttt{individuals} and \texttt{individuals\_merge} show only the middle marker filled and \texttt{sifting} show only the rightmost marker filled, indicating that optimal performance in this region requires fixed assignment to SSD for \texttt{individuals} and \texttt{individuals\_merge} and BeeGFS for \texttt{sifting}. 
\COMPRESS{The complementary insights from \Cref{fig:region-clusters} (quantitative performance separation) and \Cref{fig:dag-panels} (semantic storage rules) demonstrate \QoSFlow's dual contribution: it not only partitions configurations into low-variance performance groups but also provides actionable explanations through stage-specific storage assignments.}
This combination delivers interpretable QoS guidance that explicitly identifies which storage decisions are performance-critical and where operational flexibility exists within a performance region.

\begin{figure}[t!]
  \centering
  \includegraphics[width=0.85\columnwidth]{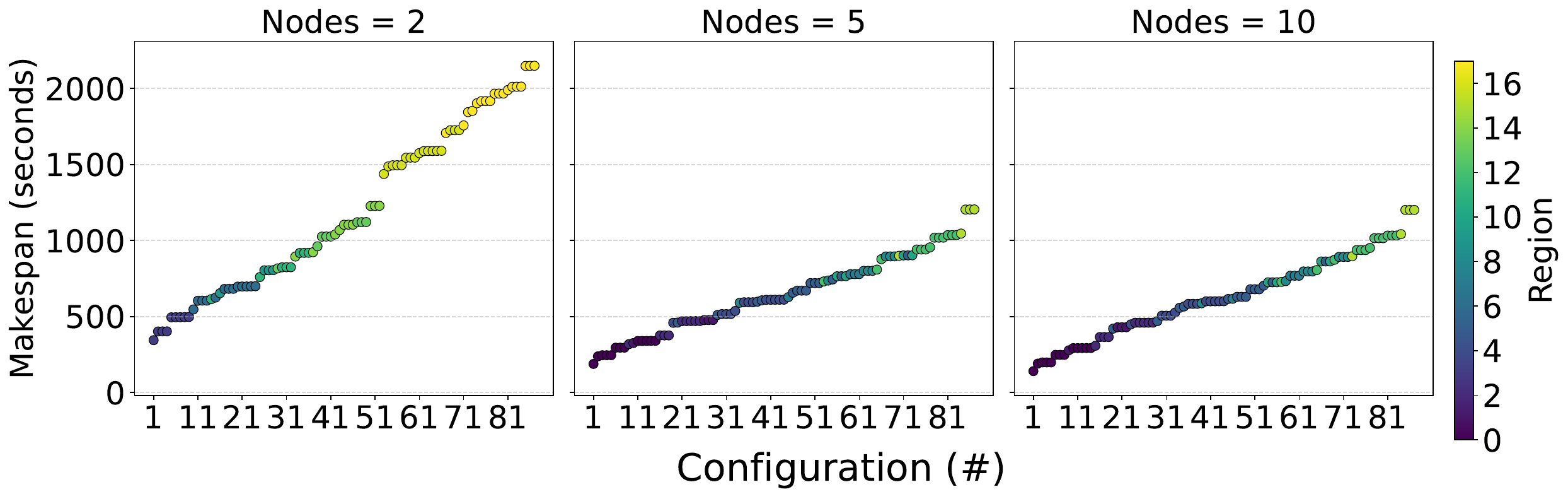}
  \caption{\texttt{1kgenome} region formation across parallelism scales ($2$, $5$, $10$ nodes).}
  \label{fig:1kg-scatter} 
 \vspace{-1em} 
\end{figure}

\COMPRESS{
{\color{blue}
\myparagraphA{Scalability through analytical projection (not additional executions)}
. In particular, when scale changes (for example, node count or input size), QoSFlow applies analytic scale projection to update stage-level I/O characteristics and DAG timing, and then re-evaluates region boundaries and critical-path shifts under the projected conditions without requiring new end-to-end workflow runs at every scale.
Importantly, this scaling does not introduce new decision dimensions of the form “which node runs what”; QoSFlow operates over stage-to-tier assignments and concurrency descriptors, so its algorithmic cost depends on the number of candidate configurations evaluated and model features, not on the number of compute nodes available.
Finally, effective scaling is bounded by workflow structure itself: if a workflow exposes limited task-level parallelism, adding nodes beyond that limit does not increase achievable throughput (for example, the final stages of \texttt{1kgenome} admit at most ten concurrent tasks).}}

\myparagraphA{Regions across parallelism scales: separation, transferability, and I/O sensitivity}
\label{sec:eval-scale-io}
From a practical deployment perspective, storage configuration decisions must retain their effectiveness as workflows scale.  
Effective scaling is bounded by workflow structure itself: if a workflow exposes limited task-level parallelism, adding nodes beyond that limit does not increase achievable throughput (for example, the final stages of \texttt{1kgenome} admit at most ten concurrent tasks).
\QoSFlow is designed to reason about larger data volumes and higher parallelism through modeling rather than direct workflow re-execution.
A critical question is whether  \QoSFlow–derived regions maintain their interpretability and performance separation when parallelism changes.
We therefore examine region formation across three distinct compute node counts.

~\Cref{fig:1kg-scatter} evaluates both separation and adaptability at $2$, $5$, and $10$ nodes. Within each panel, the \emph{ordered} configurations are colored by the region assigned by \QoSFlow at that scale, so horizontal ``bands'' indicate low within-region variance, while vertical steps between adjacent bands indicate between-region separation. The analysis demonstrates \QoSFlow consistently produces tight, well-separated bands across all scales, demonstrating that interpretable partitions persist regardless of scale. However, region memberships and even the total number of regions vary with scale, underscoring that regions are inherently scale-dependent rather than universally transferable.

\begin{figure}[t]
  \centering
  \includegraphics[width=0.8\columnwidth]{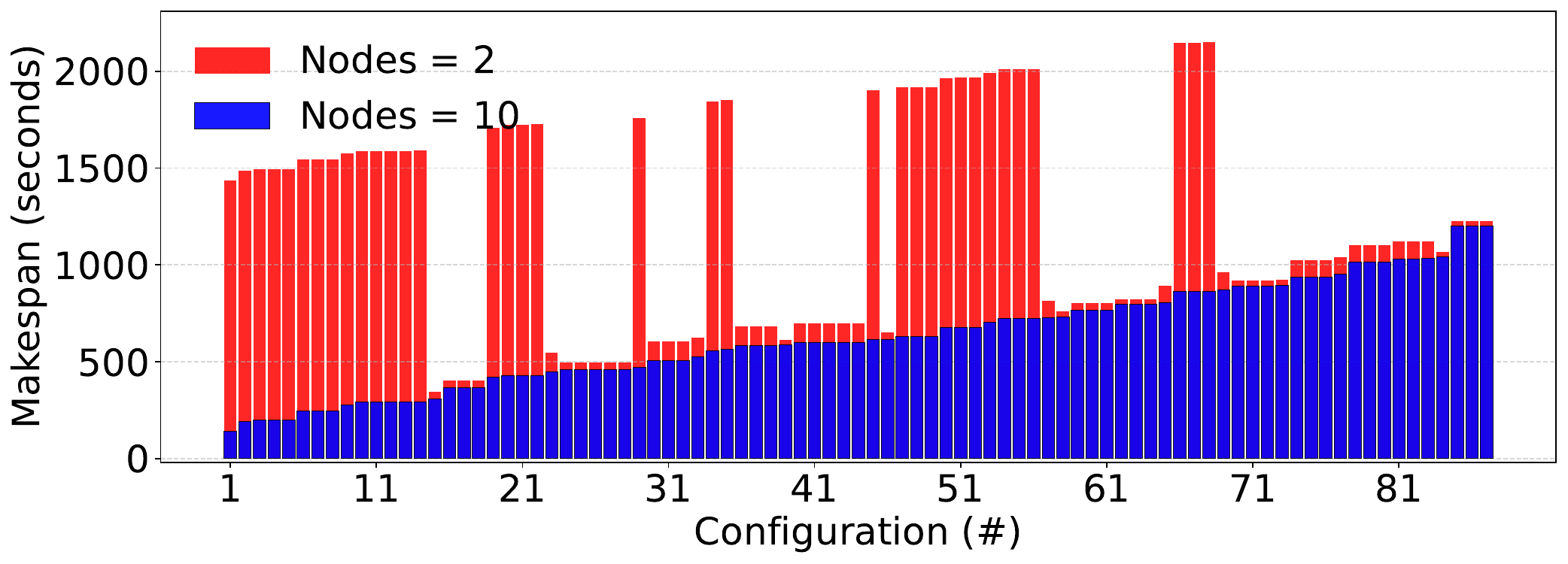}
  \caption{\texttt{1kgenome} per-configuration makespan with $10$ (blue) compute nodes overlaid on makespan with $2$ compute nodes (red).}
  \label{fig:overlay-n2-n10}
  \vspace{-1em}
\end{figure}

\begin{figure}[t!]
  \centering
  \includegraphics[width=0.85\columnwidth]{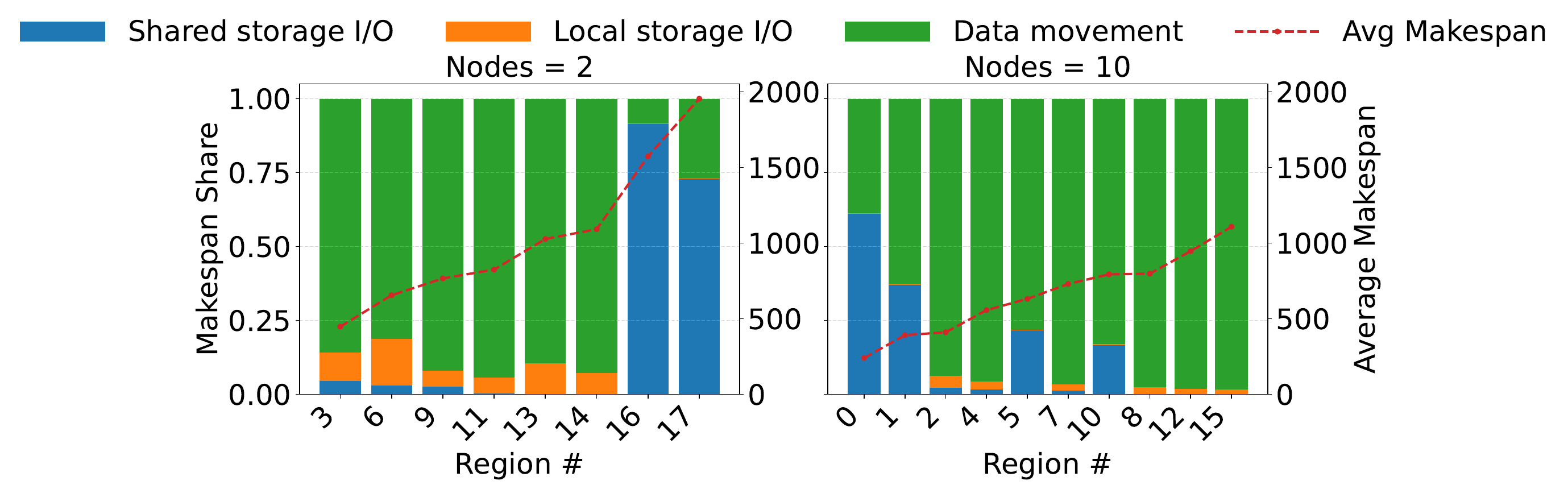}
  \caption{\texttt{1kgenome} region-level cost composition at $2$ and $10$ nodes.}
  \label{fig:1kg-cost-share}
  \vspace{-1.5em}
\end{figure}

Figure~\ref{fig:overlay-n2-n10} directly examines whether performance rankings learned at one scale transfer effectively to another. 
The visualization overlays the makespan of identical configurations at $10$ nodes (blue) against their performance at $2$ nodes (red). 
This allows crossovers and inversions to be visually exposed: configurations that achieve top rankings at 10 nodes frequently fall to middle or bottom positions at 2 nodes, and vice versa.
These pronounced performance reversals reveal non-monotonic scaling behavior, indicating that the dominant performance  bottleneck shifts with scale. Configurations optimized for smaller scales may become suboptimal at larger scales due to different I/O patterns, communication overhead, or resource utilization characteristics. 
This analysis demonstrates that assuming a single global ranking across scales is misleading and potentially counterproductive. Instead, regions should either be formed independently for each target scale, or scaling parameters must be explicitly incorporated into the feature space and model selection process.

\begin{tcolorbox}[colback=bg,left=1ex,top=1ex,boxsep=0ex,bottom=1ex,width=\linewidth]
\obs{} Region membership and ordering are not guaranteed to be transferred across different parallelism scales; regions should be formed and validated per scale.
\end{tcolorbox}

Figure~\ref{fig:1kg-cost-share} provides insight into  \emph{why} regions exhibit different performance characteristics by decomposing the critical-path cost into its constituent components.
For each region at $2$ and $10$ nodes, the stacked bars represent the proportional makespan contributions from three distinct cost sources: shared storage I/O (I/O to BeeGFS), local storage I/O (I/O to SSD/tmpFS), and data movement (data stage-in/out). The overlaid dashed line indicates the region’s median makespan, establishing the connection between cost composition and overall performance. We are interested in whether top regions are dominated by the same proportion of shares across scales and how those shares evolve. The Figure shows that high-performing regions at node scale $2$ can be limited predominantly by data movement, while at node scale $10$ execution shares (shared or local) become important; conversely, some regions retain a movement-dominated profile but with different absolute makespan. This indicates varying I/O sensitivity across scales and motivate scale-aware storage configuration assignment: the same storage configuration assignment can hit different bottlenecks as the system grows, and \QoSFlow’s regions provide a compact, actionable partition for diagnosing and prioritizing those effects.

\begin{tcolorbox}[colback=bg,left=1ex,top=1ex,boxsep=0ex,bottom=1ex,width=\linewidth]
\obs{} The dominant cost component and effective critical path can shift with scale and storage configuration; \QoSFlow regions make these shifts visible. 
\end{tcolorbox}


\begin{figure}[t!]
  \centering
  \includegraphics[width=0.85\columnwidth]{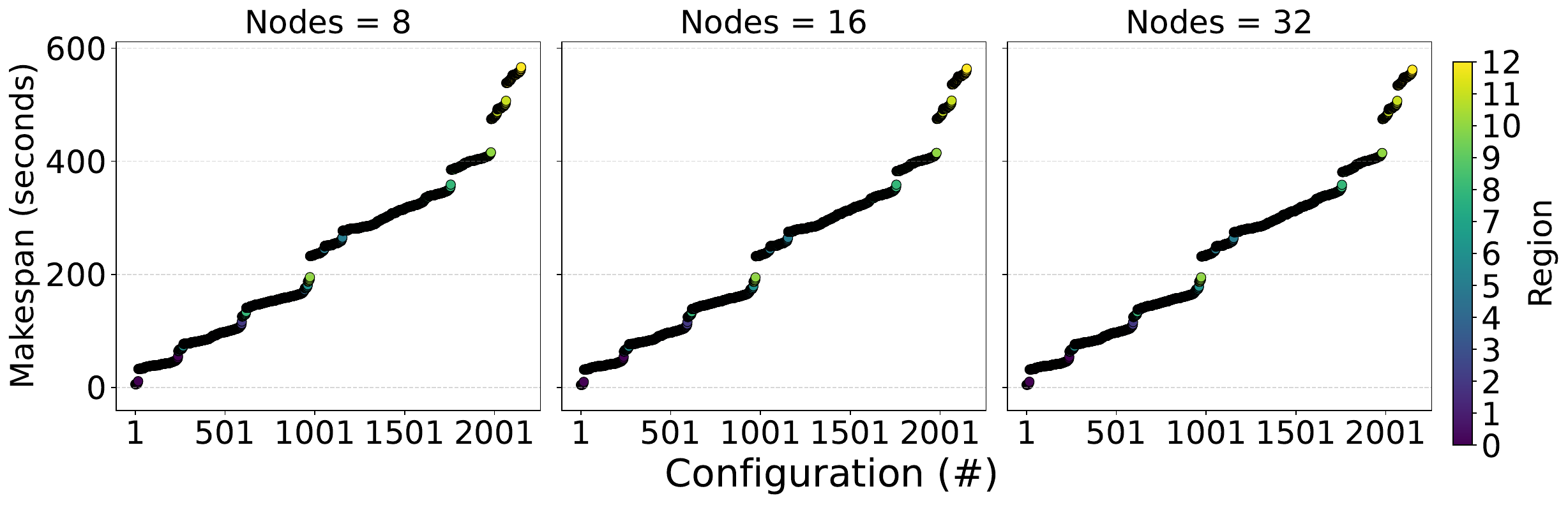}
  \caption{\QoSFlow-identified  regions for \texttt{PyFLEXTRKR} across parallelism scales ($8$, $16$, $32$ nodes).}
  \label{fig:pyflex-scatter}
  \vspace{-1em}
\end{figure}

\begin{figure}[t]
  \centering
  \includegraphics[width=0.85\columnwidth]{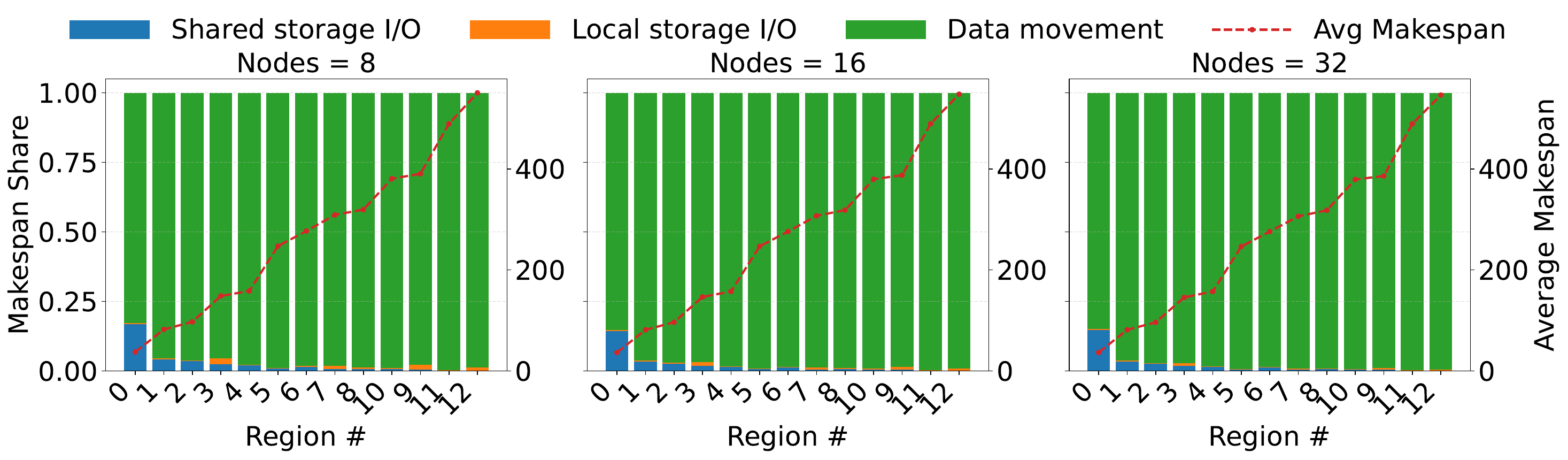}
  \caption{\texttt{PyFLEXTRKR} region-level cost composition at $8$, $16$, and $32$ nodes.}
  \label{fig:pyflex-cost-share}
  \vspace{-1.5em}
\end{figure}

\subsection{Case Study: PyFLEXTRKR}
\label{sec:wf-pyflex}
PyFLEXTRKR~\cite{feng2023pyflextrkr, pyflextrkr_git} is an atmospheric science analysis 
workflow 
which consists of nine sequential stages (the DAG in~\cref{fig:three-dags}c).
The early stages perform feature identification and mapping on gridded sensor data, and later stages compute statistics and products. 
\COMPRESS{Tasks within a stage execute in parallel, but stage-to-stage dependencies are linear across the nine-stage chain.}

\myparagraphA{Regions across parallelism scales}
Following the same analytical approach as~\Cref{fig:1kg-scatter}, ~\Cref{fig:pyflex-scatter} presents \emph{ordered} configurations colored by their \QoSFlow regions; the same band/step interpretation applies. In contrast to \texttt{1kgenome}, \texttt{PyFLEXTRKR} exhibits \emph{stable} segmentation across $8$, $16$, and $32$ nodes: the number of regions, the ordering of regions, and the per-configuration region labels are effectively unchanged across these scales, with performance bands maintaining  one-to-one alignment. This stability indicates that, the dominant critical-path structure and I/O-to-compute balance remain consistent across this node range. Consequently, \QoSFlow’s region labels and their associated scheduling guidance transfer reliably across different node counts.

\myparagraphA{Regions cost composition}
Analogous to Fig.~\ref{fig:1kg-cost-share}, Fig.~\ref{fig:pyflex-cost-share} decomposes each region’s \emph{critical-path} cost into shared storage I/O, local storage I/O, and data movement and overlays the region’s average makespan; interpretation follows \S\ref{sec:eval-scale-io}. Unlike \texttt{1kgenome}, \texttt{PyFLEXTRKR}’s region-wise composition is nearly invariant across $8$, $16$, and $32$ nodes: the same regions exhibit essentially the same shares, and the relative ordering of regions by median makespan persists. This stability suggests that the workflow’s I/O sensitivity profile does not materially shift with scale in this range, reinforcing the transferability of region-based placements for \texttt{PyFLEXTRKR}. 


\begin{figure}[t!]
  \centering
  \includegraphics[width=0.85\columnwidth]{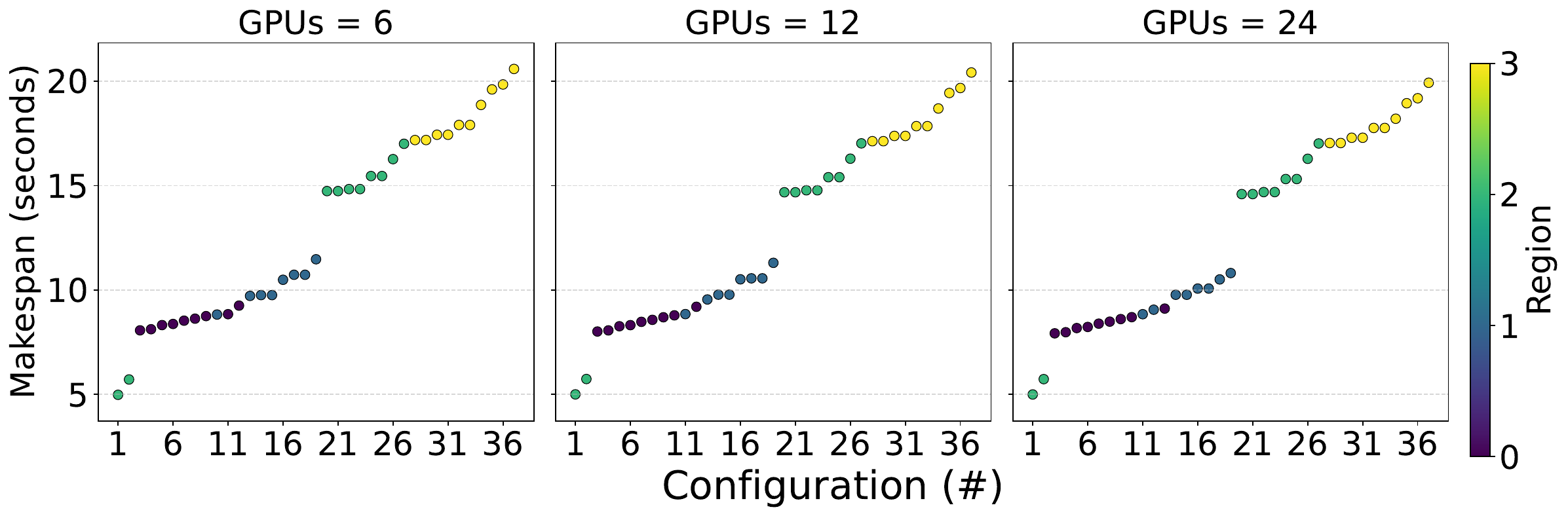}
  \caption{\QoSFlow-identified  regions for \texttt{DDMD} across 3 GPU scales($6$,$12$,$24$).} 
  \label{fig:ddmd-scatter}
  \vspace{-1em}
\end{figure}

\subsection{Case Study: DeepDriveMD (DDMD)}
\label{sec:wf-ddmd}
DeepDriveMD~\cite{lee2019deepdrivemd,ddmd_git} is an ML-steered molecular dynamics workflow that iterates a four-stage loop (the DAG in Fig.~\ref{fig:three-dags}b): parallel \textit{simulation} tasks produce trajectory/feature files, an \textit{aggregation} stage consolidates outputs, \textit{training} updates a model, and \textit{inference} scores new structures to seed the next iteration. The pipeline is iterative with many simulation tasks per iteration and single-task downstream stages. 

\begin{figure}[t!]
  \centering
  \includegraphics[width=0.85\columnwidth]{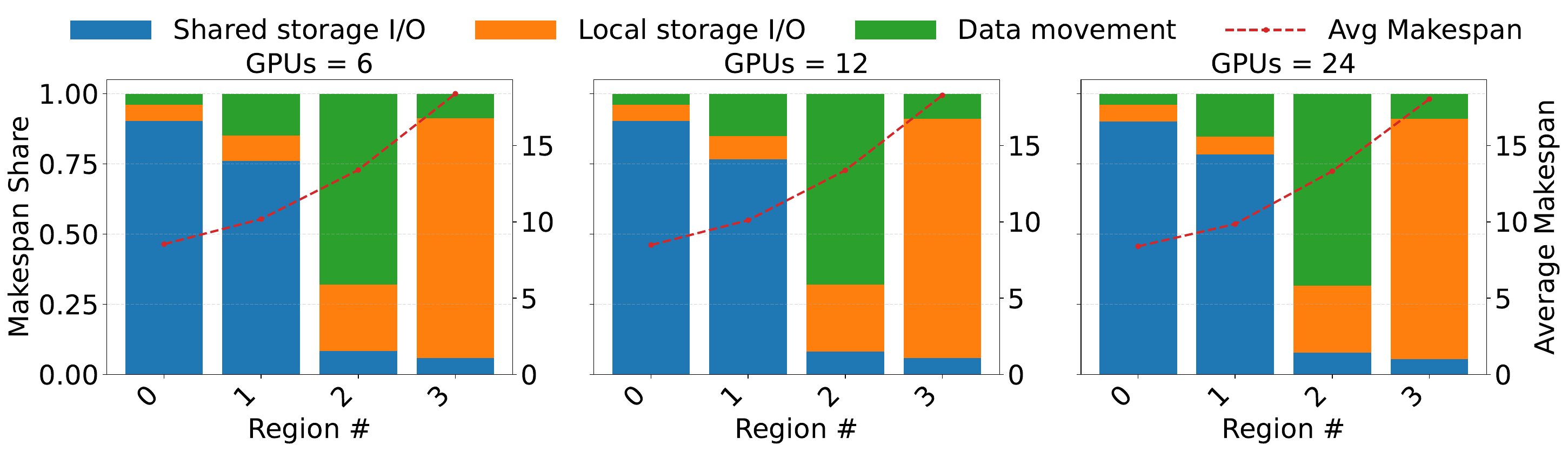}
  \caption{DDMD region-level cost composition at $6$, $12$, and $24$ GPUs.}
  \label{fig:ddmd-cost-share}
  \vspace{-1.5em}
\end{figure}

\myparagraphA{Regions across parallelism scales}
\COMPRESS{To assess how \QoSFlow segmentation performs for DDMD, we examine region formation at three node counts ($1$, $2$, $4$). ~\Cref{fig:ddmd-scatter} presents \emph{ordered} configurations colored by the regions learned at each scale; bands indicate low within-region variance and steps indicate separation. Consistent with \texttt{PyFLEXTRKR},} 
\texttt{DDMD} exhibits \emph{stable} segmentation across three scales ($1$, $2$, and $4$ nodes) (\Cref{fig:ddmd-scatter}): the number of regions, their ordering, and per-configuration labels remain effectively unchanged. This stability suggests that \texttt{DDMD}’s critical-path structure and I/O/compute balance are preserved over this range.
\COMPRESS{ enabling reliable transfer of region labels and their associated scheduling guidance across different node counts.}

\myparagraphA{Regions cost composition}
As can be observed from~\Cref{fig:ddmd-cost-share}, the region-wise cost shares are nearly invariant across $6$, $12$, and $24$ GPUs: the same regions exhibit essentially the same composition, and the relative ordering by median makespan persists. This invariance indicates that \texttt{DDMD}’s I/O sensitivity profile is stable over these scales, reinforcing the transferability of region-based storage placements.

\begin{tcolorbox}[colback=bg,left=1ex,top=1ex,boxsep=0ex,bottom=1ex,width=\linewidth]
\obs{} Different workflows exhibit different region stability; some remain consistent across scales, while others reorder markedly as critical paths shift.
\end{tcolorbox}

{
\begin{table}[t]
  \centering
  \caption{\small QoS queries validated for \QoSFlow \\ (\cmark\ MATCHED, \xmark\ No valid config.).}
  \label{tab:qos-coverage}
  \small
  \setlength{\tabcolsep}{6pt}
  \begin{tabular}{lcccc}
    \hline
    \textbf{Workflow} & \textbf{Q1} & \textbf{Q2} & \textbf{Q3} & \textbf{Q4} \\
    \hline
    1kgenome    & \cmark & \cmark & \xmark & \cmark \\
    PyFLEXTRKR  & \cmark & \cmark & \xmark & \cmark \\
    DeepDriveMD & \cmark & \cmark & \xmark & \cmark \\
    \hline
  \end{tabular}
  \vspace{-1em}
\end{table}
}  

\subsection{\QoSFlow Validation under Different QoS Requests
}
\label{sec:qos-coverage}
Operators require actionable scheduling decisions under explicit constraints rather than unconstrained optimization.  Typical requests include choosing optimal compute resource count under a resource constraint, selecting among allowable storage tiers, avoiding a particular tier, or finding the best fallback when a preferred tier is offline. The four queries below (Q1--Q4) capture this subset of operator needs in a form that \QoSFlow can evaluate against its critical-path makespan model (\Cref{subsec:enumerate-critical}) and region ranking (\Cref{subsec:region-clustering}). We evaluate these QoS requests by validating \QoSFlow's recommendations against measured execution outcomes.
\begin{enumerate}[label=Q\arabic*.]
  \item Select optimal configuration for node scaling under capacity constraints
  \item Choose best storage configuration from allowed tier subsets
  \item Find optimal storage configuration to meet a deadline while excluding specific tiers
  \item Determine best alternative configuration(s) when preferred tiers are unavailable
\end{enumerate}

\QoSFlow answers Q1 by evaluating and ranking feasible node counts with the same configuration set and selecting the minimum predicted makespan at the constrained scale (\Cref{subsec:enumerate-critical}). Q2 is supported by restricting stage--tier choices to the permitted set and ranking the resulting assignments or regions (\Cref{subsec:region-clustering}). Q3 is \emph{not} supported: if a tier is essential for 
meeting a deadline, then the QoS request is denied.
Q4 operates by excluding the unavailable tier from the feasible configuration space  and re-ranking; the critical-path and regional summaries update accordingly to recommend the best alternative. As shown in~\Cref{tab:qos-coverage}, in all cases, \QoSFlow recommendations match empirical execution results.

\section{Related work}
\label{sec:related-work}

\myparagraph{Application sensitivity analysis} 
\COMPRESS{where they define network latency sensitivity as $\lambda_L = \frac{\partial T}{\partial L}$ (the partial derivative of runtime $T$ with respect to network latency $L$). The authors convert MPI execution graphs into linear programs and leverage the reduced cost concept from linear programming to compute sensitivity measures without requiring exhaustive parameter sweeps.}
In~\cite{shen2024llamp} and~\cite{shen2025edan}, sensitivity analysis focuses on tolerance of applications to network and memory latency.
\COMPRESS{, defining absolute memory latency sensitivity $\lambda$ as $\frac{\partial T}{\partial \alpha}$ (where $\alpha$ is memory access latency).} Both generate execution graphs from application traces - LLAMP uses MPI traces with the LogGPS model, while EDAN uses instruction-level traces to create execution DAGs that expose memory-level parallelism. While these tools can predict how network and memory latency affect a single application, they cannot reason about how these effects propagate through \textit{workflows} and fail to provide QoS guidance.
\COMPRESS{The sensitivity metrics enable architects and developers to rank applications by their tolerance to latency increases, identify performance bottlenecks, and make informed decisions about system design and resource allocation.}

\myparagraph{QoS for scientific workflows} In~\cite{yu2006scheduling}, the authors addressed the problem of scheduling interdependent workflow tasks on utility grid environments while meeting user QoS requirements (deadline and budget constraints) through genetic algorithms. 
Complementary to our work, Quality of Resilience (QoR) by Tolosana-Calasanz et al.~\cite{tolosana2011characterizing} extends QoS concepts to focus on fault tolerance and reliability, proposing multi-level metrics to assess how resilient workflow enactment is likely to be in the presence of failures. 
For scientific workflows in clouds, studies like Partial Critical Path family (PCP/IC-PCP) map tasks to meet user deadlines while controlling cost~\cite{abrishami2013deadline}. 
In IaaS settings~\cite{malawski2015algorithms}, scheduling with elastic resource provisioning for workflow ensembles under cost and deadline constraints was demonstrated. In all these prior approaches for QoS, the critical path is \textbf{compute-centric} and overlooks critical performance factors like storage tier heterogeneity, I/O access patterns, and data flow characteristics, whereas \QoSFlow addresses these factors through an \textbf{I/O-centric} approach essential for modern data-intensive workflows.

\myparagraph{Workflow performance profiling}
In~\cite{krol2016workflow}, the authors focus on empirical data collection to build time-series profiles that capture how resource consumption (CPU, memory, I/O) varies over time during workflow execution and use sensitivity analysis to understand parameter impacts on job performance. 
This approach results in high computational overhead to collect data and does not provide predictive modeling capabilities or interpretable configuration space partitioning that enable rapid QoS-driven scheduling decisions offered by \QoSFlow.



\section{Conclusions}
\label{sec:conclusions}
This paper presents \QoSFlow, an interpretable quality-of-service modeling method for workflow performance analysis that addresses the critical challenge of understanding how storage and compute configurations impact execution time in HPC environments. \QoSFlow introduces a novel three-phase sensitivity-driven modeling approach that combines analytical workflow modeling with CART-based region clustering to create interpretable performance regions. The regions explain which configuration settings are critical versus flexible and why specific assignments achieve better performance. We validated the interpretable model across three scientific workflows,  demonstrating that \QoSFlow consistently creates well-separated, low-variance performance groups that outperform the best performing heuristic by 27.38\% while providing semantic explanations of region rules and cost composition.



\section*{Acknowledgments}
We sincerely thank the reviewers for their valuable feedback. This work was supported in part by the National Science Foundation (NSF) under grants CNS-2008265 and CCF-2412345.
This research is also supported by the U.S.\@ Department of Energy (DOE) through the Office of Advanced Scientific Computing Research's ``Orchestration for Distributed \& Data-Intensive Scientific Exploration'' and the ``Decentralized data mesh for autonomous materials synthesis'' AT SCALE LDRD at Pacific Northwest National Laboratory.
PNNL 
is operated by Battelle for the DOE under Contract DE-AC05-76RL01830. The authors acknowledge the assistance of ChatGPT and Claude (Anthropic) in performing editorial revisions to the manuscript.






\bibliographystyle{unsrtnat}
{
\bibliography{refs}

\begin{thebibliography}{51}
\providecommand{\natexlab}[1]{#1}
\providecommand{\url}[1]{\texttt{#1}}
\expandafter\ifx\csname urlstyle\endcsname\relax
  \providecommand{\doi}[1]{doi: #1}\else
  \providecommand{\doi}{doi: \begingroup \urlstyle{rm}\Url}\fi

\bibitem[{National Academies of Sciences, Engineering and Medicine}(2022)]{nas:workflows:2022}
{National Academies of Sciences, Engineering and Medicine}.
\newblock \emph{Automated Research Workflows For Accelerated Discovery: Closing the Knowledge Discovery Loop}.
\newblock The National Academies Press, Washington, DC, 2022.
\newblock \doi{10.17226/26532}.

\bibitem[Ferreira~da Silva et~al.(2024)Ferreira~da Silva, Badia, Bard, Foster, Jha, and Suter]{Ferreira:2024:Computer-workflow-frontiers}
Rafael Ferreira~da Silva, Rosa~M. Badia, Deborah Bard, Ian~T. Foster, Shantenu Jha, and Fr{\'e}d{\'e}ric Suter.
\newblock Frontiers in scientific workflows: Pervasive integration with high-performance computing.
\newblock \emph{Computer}, 57\penalty0 (8):\penalty0 36--44, 2024.
\newblock \doi{10.1109/MC.2024.3401542}.

\bibitem[Rashid and Dai(2025)]{rashid2025adaptbf}
Md~Hasanur Rashid and Dong Dai.
\newblock Adaptbf: Decentralized bandwidth control via adaptive token borrowing for hpc storage.
\newblock In \emph{2025 IEEE International Parallel and Distributed Processing Symposium (IPDPS)}, pages 775--788. IEEE, 2025.

\bibitem[Topcuoglu et~al.(2002{\natexlab{a}})Topcuoglu, Hariri, and Wu]{Topcuoglu:2002:scheduling}
H.~Topcuoglu, S.~Hariri, and Min-You Wu.
\newblock Performance-effective and low-complexity task scheduling for heterogeneous computing.
\newblock \emph{IEEE Transactions on Parallel and Distributed Systems}, 13\penalty0 (3):\penalty0 260--274, 2002{\natexlab{a}}.
\newblock \doi{10.1109/71.993206}.

\bibitem[Li et~al.(2022)Li, Liu, Guo, Chen, Cheng, Zheng, and Guo]{li2022faasflow}
Zijun Li, Yushi Liu, Linsong Guo, Quan Chen, Jiagan Cheng, Wenli Zheng, and Minyi Guo.
\newblock Faasflow: Enable efficient workflow execution for function-as-a-service.
\newblock In \emph{Proceedings of the 27th ACM Intl. Conf. on architectural support for programming languages and operating systems}, pages 782--796, 2022.

\bibitem[Chowdhury et~al.(2022)Chowdhury, Di~Natale, Moody, Mohror, and Yu]{Chowdhury:2022:IPDPS-dfman}
Fahim Chowdhury, Francesco Di~Natale, Adam Moody, Kathryn Mohror, and Weikuan Yu.
\newblock Dfman: A graph-based optimization of dataflow scheduling on high-performance computing systems.
\newblock In \emph{2022 IEEE Intl. Parallel and Distributed Processing Symposium (IPDPS)}, pages 368--378, 2022.
\newblock \doi{10.1109/IPDPS53621.2022.00043}.

\bibitem[Ahn et~al.(2022)Ahn, Zhang, Mast, Herbein, Di~Natale, Kirshner, Jacobs, Karlin, Milroy, De~Supinski, Van~Essen, Allen, and Lightstone]{Ahn:2022:flux-runtime-feedback}
Dong~H. Ahn, Xiaohua Zhang, Jeffrey Mast, Stephen Herbein, Francesco Di~Natale, Dan Kirshner, Sam~Ade Jacobs, Ian Karlin, Daniel~J. Milroy, Bronis De~Supinski, Brian Van~Essen, Jonathan Allen, and Felice~C. Lightstone.
\newblock Scalable composition and analysis techniques for massive scientific workflows.
\newblock In \emph{2022 IEEE 18th Intl. Conf. on e-Science (e-Science)}, pages 32--43, 2022.

\bibitem[Mayer et~al.(2017)Mayer, Mayer, and Laich]{Mayer:2017:tensorflow-critical-path}
Ruben Mayer, Christian Mayer, and Larissa Laich.
\newblock The tensorflow partitioning and scheduling problem: It's the critical path!
\newblock In \emph{Proceedings of the 1st Workshop on Distributed Infrastructures for Deep Learning}, DIDL '17, pages 1--6, New York, NY, USA, 2017. Association for Computing Machinery.
\newblock ISBN 9781450351690.
\newblock \doi{10.1145/3154842.3154843}.
\newblock URL \url{https://doi.org/10.1145/3154842.3154843}.

\bibitem[Wu et~al.(2023)Wu, Xiao, Lin, Tang, and Li]{Wu:2023:path-metis}
Baixuan Wu, Zheng Xiao, Peiying Lin, Zhuo Tang, and Kenli Li.
\newblock Critical path awareness techniques for large-scale graph partitioning.
\newblock \emph{IEEE Transactions on Sustainable Computing}, 8\penalty0 (3):\penalty0 412--422, 2023.
\newblock \doi{10.1109/TSUSC.2023.3263172}.

\bibitem[Buttazzo et~al.(2011)Buttazzo, Bini, and Wu]{Buttazzo:partition-real-time}
Giorgio Buttazzo, Enrico Bini, and Yifan Wu.
\newblock Partitioning real-time applications over multicore reservations.
\newblock \emph{IEEE Transactions on Industrial Informatics}, 7\penalty0 (2):\penalty0 302--315, 2011.
\newblock \doi{10.1109/TII.2011.2123902}.

\bibitem[Herbein et~al.(2016)Herbein, Ahn, Lipari, Scogland, Stearman, Grondona, Garlick, Springmeyer, and Taufer]{Herbein:2016:io-aware-scheduling}
Stephen Herbein, Dong~H. Ahn, Don Lipari, Thomas~R.W. Scogland, Marc Stearman, Mark Grondona, Jim Garlick, Becky Springmeyer, and Michela Taufer.
\newblock Scalable i/o-aware job scheduling for burst buffer enabled hpc clusters.
\newblock In \emph{Proceedings of the 25th ACM Intl. Symposium on High-Performance Parallel and Distributed Computing}, HPDC '16, pages 69--80, New York, NY, USA, 2016. Association for Computing Machinery.
\newblock ISBN 9781450343145.
\newblock \doi{10.1145/2907294.2907316}.
\newblock URL \url{https://doi.org/10.1145/2907294.2907316}.

\bibitem[Fauzia et~al.(2013)Fauzia, Elango, Ravishankar, Ramanujam, Rastello, Rountev, Pouchet, and Sadayappan]{Fauzia:2013:tiling-data-locality}
Naznin Fauzia, Venmugil Elango, Mahesh Ravishankar, J.~Ramanujam, Fabrice Rastello, Atanas Rountev, Louis-No\"{e}l Pouchet, and P.~Sadayappan.
\newblock Beyond reuse distance analysis: Dynamic analysis for characterization of data locality potential.
\newblock \emph{ACM Trans. Archit. Code Optim.}, 10\penalty0 (4), dec 2013.
\newblock ISSN 1544-3566.
\newblock \doi{10.1145/2541228.2555309}.
\newblock URL \url{https://doi.org/10.1145/2541228.2555309}.

\bibitem[Tang et~al.(2016)Tang, Liu, Ammar, Li, and Li]{tang2016optimized}
Zhuo Tang, Min Liu, Almoalmi Ammar, Kenli Li, and Keqin Li.
\newblock An optimized mapreduce workflow scheduling algorithm for heterogeneous computing.
\newblock \emph{The Journal of Supercomputing}, 72:\penalty0 2059--2079, 2016.

\bibitem[Wang et~al.(2014)Wang, Zhou, Li, Zhao, Lang, and Raicu]{wang2014optimizing}
Ke~Wang, Xraobing Zhou, Tonglin Li, Dongfang Zhao, Michael Lang, and Ioan Raicu.
\newblock Optimizing load balancing and data-locality with data-aware scheduling.
\newblock In \emph{2014 IEEE Intl. Conf. on Big Data (Big Data)}, pages 119--128. IEEE, 2014.

\bibitem[Firoz et~al.(2025)Firoz, Lee, Guo, Tang, Tallent, and Peng]{Firoz+:2025:SSDBM-fastflow}
Jesun Firoz, Hyungro Lee, Luanzheng Guo, Meng Tang, Nathan~R. Tallent, and Zhen Peng.
\newblock Fastflow: Rapid workflow response by prioritizing critical data flows and their interactions.
\newblock In \emph{Proc. of the 37th Intl. Conf. on Scalable Scientific Data Management}. ACM, June 2025.
\newblock \doi{10.1145/3733723.3733735}.

\bibitem[Lee et~al.(2025{\natexlab{a}})Lee, Firoz, Tallent, Guo, and Halappanavar]{Lee+:2025:IPDPS-flowforecaster}
Hyungro Lee, Jesun Firoz, Nathan~R. Tallent, Luanzheng Guo, and Mahantesh Halappanavar.
\newblock {FlowForecaster}: Automatically inferring detailed \& interpretable workflow scaling models for forecasts.
\newblock In \emph{Proc. of the 39th IEEE Intl. Parallel and Distributed Processing Symp.}, pages 420--432. IEEE Computer Society, June 2025{\natexlab{a}}.
\newblock \doi{10.1109/IPDPS64566.2025.00045}.

\bibitem[Yu and Buyya(2006)]{yu2006scheduling}
Jia Yu and Rajkumar Buyya.
\newblock Scheduling scientific workflow applications with deadline and budget constraints using genetic algorithms.
\newblock \emph{Scientific Programming}, 14\penalty0 (3-4):\penalty0 217--230, 2006.

\bibitem[Tolosana-Calasanz et~al.(2011)Tolosana-Calasanz, Lackovic, F.~Rana, Ba{\~n}ares, and Talia]{tolosana2011characterizing}
Rafael Tolosana-Calasanz, Marco Lackovic, Omer F.~Rana, Jos{\'e}~{\'A} Ba{\~n}ares, and Domenico Talia.
\newblock Characterizing quality of resilience in scientific workflows.
\newblock In \emph{Proceedings of the 6th workshop on Workflows in support of large-scale science}, pages 117--126, 2011.

\bibitem[Shen et~al.(2024)Shen, Huang, Chrapek, Schneider, Dayal, Gajbe, Wisniewski, and Hoefler]{shen2024llamp}
Siyuan Shen, Langwen Huang, Marcin Chrapek, Timo Schneider, Jai Dayal, Manisha Gajbe, Robert Wisniewski, and Torsten Hoefler.
\newblock Llamp: Assessing network latency tolerance of hpc applications with linear programming.
\newblock In \emph{SC24: International Conference for High Performance Computing, Networking, Storage and Analysis}, pages 1--18. IEEE, 2024.

\bibitem[Shen et~al.(2025)Shen, Khalilov, Gianinazzi, Schneider, Chrapek, Dayal, Gajbe, Wisniewski, and Hoefler]{shen2025edan}
Siyuan Shen, Mikhail Khalilov, Lukas Gianinazzi, Timo Schneider, Marcin Chrapek, Jai Dayal, Manisha Gajbe, Robert Wisniewski, and Torsten Hoefler.
\newblock Edan: Towards understanding memory parallelism and latency sensitivity in hpc.
\newblock In \emph{Proceedings of the 39th ACM International Conference on Supercomputing}, pages 1005--1019, 2025.

\bibitem[Kr{\'o}l et~al.(2016)Kr{\'o}l, da~Silva, Deelman, and Lynch]{krol2016workflow}
Dariusz Kr{\'o}l, Rafael~Ferreira da~Silva, Ewa Deelman, and Vickie~E Lynch.
\newblock Workflow performance profiles: development and analysis.
\newblock In \emph{European Conference on Parallel Processing}, pages 108--120. Springer, 2016.

\bibitem[Abrishami et~al.(2013)Abrishami, Naghibzadeh, and Epema]{abrishami2013deadline}
Saeid Abrishami, Mahmoud Naghibzadeh, and Dick~HJ Epema.
\newblock Deadline-constrained workflow scheduling algorithms for infrastructure as a service clouds.
\newblock \emph{Future generation computer systems}, 29\penalty0 (1):\penalty0 158--169, 2013.

\bibitem[Malawski et~al.(2015)Malawski, Juve, Deelman, and Nabrzyski]{malawski2015algorithms}
Maciej Malawski, Gideon Juve, Ewa Deelman, and Jarek Nabrzyski.
\newblock Algorithms for cost-and deadline-constrained provisioning for scientific workflow ensembles in iaas clouds.
\newblock \emph{Future Generation Computer Systems}, 48:\penalty0 1--18, 2015.

\bibitem[Rashid et~al.(2025)Rashid, Li, He, Bao, and Dai]{rashid2025dial}
Md~Hasanur Rashid, Xinyi Li, Youbiao He, Forrest~Sheng Bao, and Dong Dai.
\newblock Dial: Decentralized i/o autotuning via learned client-side local metrics for parallel file system.
\newblock In \emph{2025 IEEE 25th International Symposium on Cluster, Cloud and Internet Computing (CCGrid)}, pages 01--04. IEEE, 2025.

\bibitem[Rashid et~al.(2023)Rashid, He, Bao, and Dai]{rashid2023iopathtune}
Md~Hasanur Rashid, Youbiao He, Forrest~Sheng Bao, and Dong Dai.
\newblock Iopathtune: Adaptive online parameter tuning for parallel file system i/o path.
\newblock \emph{arXiv preprint arXiv:2301.06622}, 2023.

\bibitem[Dong et~al.(2025)Dong, Rashid, Xu, and Dai]{dong2025rl4sys}
Jiaxin Dong, Md~Hasanur Rashid, Helen Xu, and Dong Dai.
\newblock Rl4sys: A lightweight system-driven rl framework for drop-in integration in system optimization.
\newblock In \emph{Proceedings of the SC'25 Workshops of the International Conference for High Performance Computing, Networking, Storage and Analysis}, pages 1406--1414, 2025.

\bibitem[Egersdoerfer et~al.(2024)Egersdoerfer, Rashid, Dai, Fang, and Nathan]{egersdoerfer2024understanding}
Chris Egersdoerfer, Md~Hasanur Rashid, Dong Dai, Bo~Fang, and Tallent Nathan.
\newblock Understanding and predicting cross-application i/o interference in hpc storage systems.
\newblock In \emph{SC24-W: Workshops of the International Conference for High Performance Computing, Networking, Storage and Analysis}, pages 1330--1339. IEEE, 2024.

\bibitem[Deelman et~al.(2015)Deelman, Vahi, Juve, Rynge, Callaghan, Maechling, Mayani, Chen, Da~Silva, Livny, et~al.]{deelman2015pegasus}
Ewa Deelman, Karan Vahi, Gideon Juve, Mats Rynge, Scott Callaghan, Philip~J Maechling, Rajiv Mayani, Weiwei Chen, Rafael~Ferreira Da~Silva, Miron Livny, et~al.
\newblock Pegasus, a workflow management system for science automation.
\newblock \emph{Future Generation Computer Systems}, 46:\penalty0 17--35, 2015.

\bibitem[Lee et~al.(2025{\natexlab{b}})Lee, Firoz, Tallent, and Guo]{FlowForecasterRepo}
Hyungro Lee, Jesun Firoz, Nathan~R. Tallent, and Luanzheng Guo.
\newblock {FlowForecaster}.
\newblock \url{https://github.com/pnnl/FlowForecaster}, 2025{\natexlab{b}}.
\newblock Accessed: [07-01-2025].

\bibitem[{Anonymized for double blind review}(2025)]{Tang+:2025:DPM}
{Anonymized for double blind review}.
\newblock {DPM}: Dataflow performance matching for effective scheduling of data-intensive workflows.
\newblock In \emph{Under Review.}, 2025.

\bibitem[Hell and Ne{\v s}et{\v r}il(1992)]{Hell:1992:graph-core}
Pavol Hell and Jaroslav Ne{\v s}et{\v r}il.
\newblock The core of a graph.
\newblock \emph{Discrete Mathematics}, 109\penalty0 (1):\penalty0 117--126, 1992.
\newblock \doi{https://doi.org/10.1016/0012-365X(92)90282-K}.

\bibitem[Loewe et~al.(2003)Loewe, McLarty, and Morrone]{ior_tool}
William Loewe, Tyce McLarty, and Christopher Morrone.
\newblock {IOR} filesystem benchmark.
\newblock \url{https://github.com/hpc/ior}, January 2003.
\newblock Accessed: 2024-1-2.

\bibitem[Da~Veiga et~al.(2021)Da~Veiga, Gamboa, Iooss, and Prieur]{da2021basics}
S{\'e}bastien Da~Veiga, Fabrice Gamboa, Bertrand Iooss, and Cl{\'e}mentine Prieur.
\newblock \emph{Basics and trends in sensitivity analysis: Theory and practice in R}.
\newblock SIAM, 2021.

\bibitem[Smith(2024)]{smith2024uncertainty}
Ralph~C Smith.
\newblock \emph{Uncertainty quantification: theory, implementation, and applications}.
\newblock SIAM, 2024.

\bibitem[Saltelli et~al.(2004)Saltelli, Tarantola, Campolongo, Ratto, et~al.]{saltelli2004sensitivity}
Andrea Saltelli, Stefano Tarantola, Francesca Campolongo, Marco Ratto, et~al.
\newblock \emph{Sensitivity analysis in practice: a guide to assessing scientific models}, volume~1.
\newblock Wiley Online Library, 2004.

\bibitem[Borgonovo and Plischke(2016)]{borgonovo2016sensitivity}
Emanuele Borgonovo and Elmar Plischke.
\newblock Sensitivity analysis: A review of recent advances.
\newblock \emph{European Journal of Operational Research}, 248\penalty0 (3):\penalty0 869--887, 2016.

\bibitem[Borgonovo(2008)]{borgonovo2008sensitivity}
Emanuele Borgonovo.
\newblock Sensitivity analysis of model output with input constraints: A generalized rationale for local methods.
\newblock \emph{Risk Analysis: An International Journal}, 28\penalty0 (3):\penalty0 667--680, 2008.

\bibitem[Pianosi et~al.(2016)Pianosi, Beven, Freer, Hall, Rougier, Stephenson, and Wagener]{pianosi2016sensitivity}
Francesca Pianosi, Keith Beven, Jim Freer, Jim~W Hall, Jonathan Rougier, David~B Stephenson, and Thorsten Wagener.
\newblock Sensitivity analysis of environmental models: A systematic review with practical workflow.
\newblock \emph{Environmental Modelling \& Software}, 79:\penalty0 214--232, 2016.

\bibitem[Breiman et~al.(2017)Breiman, Friedman, Olshen, and Stone]{breiman2017classification}
Leo Breiman, Jerome Friedman, Richard~A Olshen, and Charles~J Stone.
\newblock \emph{Classification and regression trees}.
\newblock Chapman and Hall/CRC, 2017.

\bibitem[Hedges(1982)]{hedges1982estimation}
Larry~V Hedges.
\newblock Estimation of effect size from a series of independent experiments.
\newblock \emph{Psychological bulletin}, 92\penalty0 (2):\penalty0 490, 1982.

\bibitem[Herold et~al.(2014)Herold, Breuner, and Heichler]{herold2014introduction}
Frank Herold, Sven Breuner, and Jan Heichler.
\newblock An introduction to beegfs.
\newblock \emph{ThinkParQ, Kaiserslautern, Germany, Tech. Rep}, 2014.

\bibitem[Clarke et~al.(2012)Clarke, Zheng-Bradley, Smith, Kulesha, Xiao, Toneva, Vaughan, Preuss, Leinonen, Shumway, et~al.]{clarke20121000}
Laura Clarke, Xiangqun Zheng-Bradley, Richard Smith, Eugene Kulesha, Chunlin Xiao, Iliana Toneva, Brendan Vaughan, Don Preuss, Rasko Leinonen, Martin Shumway, et~al.
\newblock The 1000 genomes project: data management and community access.
\newblock \emph{Nature methods}, 2012.

\bibitem[1kG()]{1kG_git}
1000genomes workflow git repo.
\newblock \url{https://github.com/pegasus-isi/1000genome-workflow}.
\newblock Accessed: 2025-10-08.

\bibitem[Pan et~al.(2018)Pan, Xiong, Shen, Wang, and Jiang]{pan2018h}
Fengfeng Pan, Jin Xiong, Yijie Shen, Tianshi Wang, and Dejun Jiang.
\newblock H-scheduler: Storage-aware task scheduling for heterogeneous-storage spark clusters.
\newblock In \emph{2018 IEEE 24th International Conference on Parallel and Distributed Systems (ICPADS)}, pages 1--9. IEEE, 2018.

\bibitem[Zaharia et~al.(2010)Zaharia, Borthakur, Sen~Sarma, Elmeleegy, Shenker, and Stoica]{zaharia2010delay}
Matei Zaharia, Dhruba Borthakur, Joydeep Sen~Sarma, Khaled Elmeleegy, Scott Shenker, and Ion Stoica.
\newblock Delay scheduling: a simple technique for achieving locality and fairness in cluster scheduling.
\newblock In \emph{Proceedings of the 5th European conference on Computer systems}, pages 265--278, 2010.

\bibitem[Topcuoglu et~al.(2002{\natexlab{b}})Topcuoglu, Hariri, and Wu]{topcuoglu2002performance}
Haluk Topcuoglu, Salim Hariri, and Min-You Wu.
\newblock Performance-effective and low-complexity task scheduling for heterogeneous computing.
\newblock \emph{IEEE transactions on parallel and distributed systems}, 13\penalty0 (3):\penalty0 260--274, 2002{\natexlab{b}}.

\bibitem[Kendall(1938)]{kendall1938new}
Maurice~G Kendall.
\newblock A new measure of rank correlation.
\newblock \emph{Biometrika}, 30\penalty0 (1-2):\penalty0 81--93, 1938.

\bibitem[Feng et~al.(2023)Feng, Hardin, Barnes, Li, Leung, Varble, and Zhang]{feng2023pyflextrkr}
Zhe Feng, Joseph Hardin, Hannah~C Barnes, Jianfeng Li, L~Ruby Leung, Adam Varble, and Zhixiao Zhang.
\newblock Pyflextrkr: A flexible feature tracking python software for convective cloud analysis.
\newblock \emph{Geoscientific Model Development}, 16\penalty0 (10):\penalty0 2753--2776, 2023.

\bibitem[pyf()]{pyflextrkr_git}
Pyflextrkr workflow git repo.
\newblock \url{https://github.com/FlexTRKR/PyFLEXTRKR}.
\newblock Accessed: 2025-10-08.

\bibitem[Lee et~al.(2019)Lee, Turilli, Jha, Bhowmik, Ma, and Ramanathan]{lee2019deepdrivemd}
Hyungro Lee, Matteo Turilli, Shantenu Jha, Debsindhu Bhowmik, Heng Ma, and Arvind Ramanathan.
\newblock Deepdrivemd: Deep-learning driven adaptive molecular simulations for protein folding.
\newblock In \emph{2019 IEEE/ACM Third Workshop on Deep Learning on Supercomputers (DLS)}, pages 12--19. IEEE, 2019.

\bibitem[ddm()]{ddmd_git}
Deepdrivemd workflow git repo.
\newblock \url{https://github.com/radical-collaboration/DeepDriveMD}.
\newblock Accessed: 2025-10-08.

\end{thebibliography}
}


\end{document}